\documentclass[12pt,preprint]{aastex61}

\shorttitle{TMTS: Overview and Performance}
\shortauthors{Zhang et al.}

\begin{document}

\title{The Tsinghua University-Ma Huateng Telescopes for Survey: Overview and Performance of the System}

\author{Ji-Cheng Zhang}
\affiliation{Physics Department and Tsinghua Center for Astrophysics, Tsinghua University, Beijing, 100084, China}

\author{Xiao-Feng Wang}
\affiliation{Physics Department and Tsinghua Center for Astrophysics, Tsinghua University, Beijing, 100084, China}

\author{Jun Mo}
\affiliation{Physics Department and Tsinghua Center for Astrophysics, Tsinghua University, Beijing, 100084, China}

\author{Gao-Bo Xi}
\affiliation{Physics Department and Tsinghua Center for Astrophysics, Tsinghua University, Beijing, 100084, China}

\author{Jie Lin}
\affiliation{Physics Department and Tsinghua Center for Astrophysics, Tsinghua University, Beijing, 100084, China}

\author{Xiao-Jun Jiang}
\affiliation{National Astronomical Observatories of China, Chinese Academy of Sciences, Beijing, 100012, China}
\affiliation{School of Astronomy and Space Science, University of Chinese Academy of Sciences, Beijing, 100049, China}

\author{Xiao-Ming Zhang}
\affiliation{National Astronomical Observatories of China, Chinese Academy of Sciences, Beijing, 100012, China}

\author{Wen-Xiong Li}
\affiliation{Physics Department and Tsinghua Center for Astrophysics, Tsinghua University, Beijing, 100084, China}

\author{Sheng-Yu Yan}
\affiliation{Physics Department and Tsinghua Center for Astrophysics, Tsinghua University, Beijing, 100084, China}

\author{Zhi-Hao Chen}
\affiliation{Physics Department and Tsinghua Center for Astrophysics, Tsinghua University, Beijing, 100084, China}

\author{Lei Hu}
\affiliation{Purple Mountain Observatory, Chinese Academy of Sciences, Nanjing 210008, China}
\affiliation{School of Astronomy and Space Sciences, University of Science and Technology of China, Hefei, 230029, China}

\author{Xue Li}
\affiliation{Physics Department and Tsinghua Center for Astrophysics, Tsinghua University, Beijing, 100084, China}

\author{Wei-Li Lin}
\affiliation{Physics Department and Tsinghua Center for Astrophysics, Tsinghua University, Beijing, 100084, China}

\author{Han Lin}
\affiliation{Physics Department and Tsinghua Center for Astrophysics, Tsinghua University, Beijing, 100084, China}

\author{Cheng Miao}
\affiliation{Physics Department and Tsinghua Center for Astrophysics, Tsinghua University, Beijing, 100084, China}

\author{Li-Ming Rui}
\affiliation{Physics Department and Tsinghua Center for Astrophysics, Tsinghua University, Beijing, 100084, China}

\author{Han-Na Sai}
\affiliation{Physics Department and Tsinghua Center for Astrophysics, Tsinghua University, Beijing, 100084, China}

\author{Dan-Feng Xiang}
\affiliation{Physics Department and Tsinghua Center for Astrophysics, Tsinghua University, Beijing, 100084, China}

\author{Xing-Han Zhang}
\affiliation{Physics Department and Tsinghua Center for Astrophysics, Tsinghua University, Beijing, 100084, China}

\begin{abstract}

Over the past decade, time-domain astronomy in optical bands has developed rapidly with the operations of some wide-field survey facilities. However, most of these surveys are conducted with only a single band, and simultaneous color information is usually unavailable for the objects monitored during the survey. Here we present introductions to the system of Tsinghua University-Ma Huateng Telescopes for Survey (TMTS), which consists of an array of four optical telescopes installed on a single equatorial mount. Such a system is designed to get multiband photometry simultaneously for stars and transients discovered during the survey. The optics of each telescope is a modified Hamilton-Newtonian system, covering the wavelengths from 400 $nm$ to 900 $nm$, with a field of view (FoV) of about 4.5 deg$^2$ and a plate scale of 1.86$''$/pixel when combining with a 4K$\times$4K QHY4040 CMOS detector. The TMTS system can have a FoV of about 9 deg$^2$ when monitoring the sky with two bands (i.e., SDSS $g$ and $r$ filters) at the same time, and a maximum FoV of $\sim$18 deg$^2$ when four telescopes monitor different sky areas in monochromatic filter mode. For an exposure time of 60s, the average 3-$\sigma$ detection limit of the TMTS system can reach at $\sim$19.4 mag in Luminous filter and at $\sim$18.7 mag in SDSS $r$ filter. The preliminary discovery obtained during the first few months' survey is briefly discussed. As this telescope array is located at the Xinglong Observatory of NAOC, it can have an excellent synergy with the spectroscopic survey by the LAMOST (with a FoV of about 20 deg$^2$) at the same site, which will benefit the studies of stellar and binary physics besides the transient sciences.

\end{abstract}


\section{INTRODUCTION}

Time-domain astronomy is a new frontier in astrophysics, opening a new parameter space to explore time-varying universe by repeating observations of the same sky area. One of the pioneering surveys is the Lick Observatory Supernova Search (LOSS; \cite{2000AIPC..522..103L}, \cite{2001ASPC..246..121F}) started in the last century, which used a narrow-field reflecting telescope with an aperture of 0.76 m and a 512$\times$512 pixel charge coupled device (CCD) chip. Although the FoV of LOSS is very small, it was designed for a survey of thousands of nearby galaxies and became the most successful nearby supernova search engine for about a decade (i.e., from 1998 to 2008)\citep{2011MNRAS.412.1441L}.

With the progresses in large-area CCD detectors and techniques of massive data processing as well as the realization of robotization and intelligentization for the observing system, many wide-field optical surveys have been inspired for time-domain astronomy over the past decade. These wide-field surveys are more efficient in discovering transients on stellar and galactic scales, like variable stars, supernovae, active galactic nuclei (AGN), and tidal disruption events (TDE), etc. The Palomar Transient Factory (PTF; \cite{2009PASP..121.1334R}, \cite{2009PASP..121.1395L}) is one of such a representative wide-field survey project, followed by the Intermediate Palomar Transient Factory (iPTF; \cite{2018ApJ...854L..13H}, \cite{2019ApJ...881..128C}) survey. Both of these two surveys can cover the sky area of 7.8 deg$^2$ with one exposure. The Zwicky Transient Facility (ZTF; \cite{2019PASP..131a8002B}), equipped with an upgraded large array of CCD detectors, provides an amazing large FoV of about 47 deg$^2$ with a pixel scale of 1.0$''$/pixel, which can monitor the whole northern visible sky with a 1-2 day cadence.

Other wide-field surveys include the Deep Lens Survey (DLS; \cite{2004ApJ...611..418B}), the Dark Energy Survey (DES; \cite{2005astro.ph.10346T}), the Equation of State: SupErNovae trace Cosmic Expansion (ESSENCE; \cite{2007ApJ...666..674M}), the Supernova Legacy Survey (SNLS; \cite{2006AA...447...31A}), the Robotic Optical Transient Search Experiment (ROTSE; \cite{2003PASP..115..132A}), the Sloan Digital Sky Survey-II Supernova (SDSS-II SN; \cite{2008AJ....135..338F}) Survey, the Catalina Real-Time Transient Survey (CRTS; \cite{2009ApJ...696..870D}, \cite{2011arXiv1102.5004D}), the Panoramic Survey Telescope And Rapid Response System (Pan-STARRS; \cite{2002SPIE.4836..154K}, \cite{2016arXiv161205560C}), the SkyMapper (\cite{2007PASA...24....1K}), the La Silla-QUEST (LSQ; \cite{2012IAUS..285..324H}, \cite{2013PASP..125..683B}), the Kiso Supernova Survey (KISS; \cite{2014PASJ...66..114M}), the All-Sky Automated Survey for Supernovae (ASAS-SN; \cite{2017PASP..129j4502K}, \cite{2019ApJ...876..115S}), the Asteroid Terrestrial-impact Last Alert System (ATLAS; \cite{2011PASP..123...58T}, \cite{2018PASP..130f4505T}), the ongoing Distance Less Than 40 Mpc (DLT40; \cite{2017ApJ...851L..48Y}, \cite{2018ApJ...853...62T}) supernova search, the Evryscope survey\citep{2019PASP..131g5001R}, the Tsinghua University-National Astronomical Observatories, Chinese Academy of Sciences (NAOC) Transient Survey (TNTS; \cite{2015RAA....15..215Z}), and the Purple Mountain Observatory-Tsinghua Supernova Survey (PTSS; W. X. Li et al. 2020 in preparation), etc. These surveys aim at different scientific objectives because of different telescope apertures, FoV, limiting magnitudes. For example, the Evryscope project, having 27 small individual optical telescopes installed on one equatorial mount, could achieve over 8000 deg$^2$ for one exposure but it can only monitor very bright stars (i.e., m$_{g}$ $<$ 16.0 mag). When the next-generation facility like the Large Synoptic Survey Telescope (LSST; \cite{2002SPIE.4836...10T}) and the Chinese Space Station Optical Survey (CSS-OS; \cite{2018MNRAS.480.2178C}) come into use, the time-domain astronomy will enter into a new era because that their detection limits are superior deep than current surveys(\cite{2019ApJ...873..111I}). Moreover, the CSS-OS will have a spacial resolution comparable to that of the Hubble Space Telescope (HST) but it has a much larger FoV in comparison with the latter.

\begin{table}[htbp]
\scriptsize
\centering
\caption{Ground-based Optical Transient Surveys Relevant with Supernova Search Around the World}
\begin{tabular}{lllllllll}
\hline
\hline
Surveys				&	Telescope	(m)			&	FoV	(deg$^2$ )&	Mag. Limit.		&	Survey Area						&	Cadence				&	Status	\\
\hline
LOSS				&	0.76 					&	0.012		&	W $\sim$19.0		&	5000 nearby galaxies				&	3$\sim$5 days			& Since 1997	\\	
DLS					&	4.0					&	4.0			&	W $\sim$24.0		& 	High galactic latitude					& 	Minutes$\sim$month		& 1999-2005	\\
ESSENCE			&	4.0					&	0.36			&	W $\sim$24.0		&	Equatorial area						&	4 days				& 2002-2008	\\
SNLS				&	3.6					&	1.0			&	$R$ $\sim$24.3		&	1300deg$^2$						& 	3 days$\sim$5 years		& 2003-2008	\\	
ROTSE-III				&	0.45					&	3.42			&	W $\sim$18.5		&	Entire sky							&	...					& Since 2003	\\	
SDSS-II SN			&	2.5					&	1.5 			&	$g$$\sim$23.2		& 	Southern stripe 82					&	2 days				& 2005-2008	\\
CRTS(CSS)			&	0.7					&	8.1			&	$V$$\sim$19.5		& +70$^{\circ}$$\ge$Dec.$\ge$-30$^{\circ}$ 	& 	10 minutes$\sim$year	& Since 2007	\\
Pan-STARRS1 3$\pi$	&	1.8					&	7.0 			&	$g$$\sim$23.3		& 	Dec.$\ge$-30$^{\circ}$ 				&	7 days				& Since 2009	\\
SkyMapper			&	1.35					&	5.7 			&	$g$$\sim$21.2		&	 Entire southern sky					&	Hours$\sim$years		& Since 2009	\\
PTF					&	1.2					&	7.9 			&	$R$ $\sim$21.0 	& 	Dec.$\ge$-30$^{\circ}$				&	1minute$\sim$5days		& 2009-2012	\\
LSQ					&	1.0					&	9.6			&	$V$$\sim$20.5		& +25$^{\circ}$$\ge$Dec.$\ge$-25$^{\circ}$ 	& 	Hours$\sim$days		& 2009-2015	\\
KISS					&	1.05					&	4.6			&	$g$$\sim$20.0-21.0	& 	SDSS fields with high SFR			& 	$\sim$1hour 			& Since 2012	\\
ASAS-SN				&	24$\times$0.14			&	4$\times$20.25 &	$g$$\sim$18.0 		&	Entire sky							&	1 day				& Since 2013	\\
iPTF					&	1.2					&	7.3 			&	$g$$\sim$21.0		&	Dec.$\ge$-30$^{\circ}$ 				&	90 seconds$\sim$5days	& 2013-2017	\\
DES					&	4.0					&	3.0			&	$i$$\sim$24.0		&	The South Galactic Cap				&	$\sim$7 days			& 2013-2019	\\
Evryscope				& 2$\times$27$\times$0.061	& 	8600+8150	&	$g$$\sim$16.0		& 	Entire sky							& 	2 minute$\sim$hours		& Since 2015	\\	
ATLAS				&	2$\times$0.5			&	2$\times$30 	&	$o$$\sim$19.5		& +90$^{\circ}$$\ge$Dec.$\ge$-45$^{\circ}$ 	& 	1$\sim$2days			& Since 2015	\\
DLT40 				&	0.41					&	0.027		&	$r$$\sim$ 19.0		& Nearby galaxies$\leq$ 40Mpc 			& 	1 day				& Since 2016	\\
ZTF					&	1.2					&	47.0 			&	$R$ $\sim$20.7		& 	Dec.$\ge$-30$^{\circ}$ 				&	3750 deg$^2$/hour		& Since 2017	\\
LSST				&	8.4					&	9.6			&	$R$ $\sim$24.5		&	Entire southern sky					&	1 minute$\sim$10 years	& Future		\\	
TNTS				& 	0.6 					&	2.3 			&	$V$$\sim$19.5		& +60$^{\circ}$$\ge$Dec.$\ge$0$^{\circ}$ 	& 	3$\sim$4 days			& Since 2012	\\
PTSS				&	1.04					&	9.0 			&	$V$$\sim$20.5 		&	Dec.$\ge$-30$^{\circ}$				&	3$\sim$5 days			& Since 2016	\\
TMTS				&	4$\times$0.4			&	4$\times$4.5	&	$W$$\sim$19.4		& +70$^{\circ}$$\ge$Dec.$\ge$10$^{\circ}$ 	& 	10 seconds$\sim$1 day	& Since 2019	\\
\hline
\end{tabular}
\label{tab:surveys}
\end{table}

Table~\ref{tab:surveys} lists the ground-based optical surveys relevant to the purpose of time-domain astronomy. We compared some of the key parameters for these surveys, including diameter of the telescopes, FoV, limiting magnitude, survey area, typical survey cadence and its corresponding status. Most of the current surveys have a survey depth $<$ 20.0 mag in $R$ band, focusing on targets in local universe; while some large-telescope surveys can reach a limited magnitude of about 25.0 mag such as LSST, could explore transients in the high-z universe. Survey cadence is another key parameter affecting the transient discovery, which depends on the FoV of the telescopes as well as the scientific objectives. Those surveys with large FoV could monitor the sky with relatively high cadences and have chances to detect the early-phase signals after explosions or short-period transients.

However, none of the existing surveys aim to detect the transients in different bands at the same time. The simultaneous color information reveals underlying mechanisms of many transients, such as novae, supernovae, counterparts of gravitational wave, variable stars and eclipse binaries. For instance, several models with different color evolution in the early time have been proposed for type Ia supernovae (SNe Ia). Of these, $''$He-detonation$''$ model has two detonations in the He-shell and carbon-oxygen core, respectively \citep{1982ApJ...253..798N,1982ApJ...257..780N,2011ApJ...734...38W}, resulting in blue-red-blue color evolution in the very early phase \citep{2018ApJ...861...78M,2019ApJ...873...84P}. On the other hand, the interaction of SN ejecta with a non-degenerate companion star \citep{2010ApJ...708.1025K} or circumstellar material ejected before supernova explosion \citep{2014MNRAS.441..532D} is expected to produce a blue-red color evolution immediately after explosion. Another example is the recently discovered kilonova, the electromagnetic counterpart of gravitational wave emission due to merger of two neutron stars \citep{2017Sci...358.1556C,2017Natur.551...64A}. The optical emission in the first few days has a blue component, which is powered by Fe-group and light $r$-process nuclei, followed by a red component produced by lanthanides-rich and heavy $r$-process nuclei \citep{1998ApJ...507L..59L,2012ApJ...746...48M}. Thus, timely multiband observations can help better constrain different models and disclose the physical properties of kilonovae. For the studies of variable stars, multiband observations at the same time allow accurate estimates of change in stellar radius, when combining with stellar evolution models. For eclipse binaries, multiband observations at the same time allow extracting the spectral characteristics of the foreground star from the superpositional photometry data requires multiband photometry simultaneously at the moment of fully eclipse.

Owing to the design of telescopes, adopted survey mode and time delay caused by data reduction, normal surveys listed in Table~\ref{tab:surveys} have difficulties in obtaining simultaneous color information for detected transients,  especially for those fast-evolving objects such as fast radio bursts (FRB), gamma-ray burst (GRB) afterglow and flare stars which show significant light/color variations within hours, minutes and even seconds. In order to get precise color and temperature evolution of detected transients at the same time, we developed the TMTS system, which can collect multiband information of the transients discovered during the survey. Furthermore, TMTS has the ability to capture rapidly-evolved transients due to the use of complementary metal-oxide-semiconductor (CMOS) detectors. The color is useful to discriminate against flares of late-type dwarfs and pick the GRB afterglow candidates \citep{2013ApJ...779...18B,2018ApJ...854L..13H}. Meanwhile, such a system can be also used for monochromatic survey to get a maximum FoV up to $\sim$18 deg$^2$ and can improve to be about $\sim$25 deg$^2$ when 6K$\times$6K detectors replaced in the future. The TMTS system is located at Xinglong Observatory of NAOC, 117$^{\circ}$34$'$39$''$ East, 40$^{\circ}$23$'$26$''$ North (see \cite{2015PASP..127.1292Z} and \cite{2016PASP..128j5004Z} for the detailed parameters of this site), where the Large Sky Area Multi-Object Fiber Spectroscopic Telescope (LAMOST; \cite{2012RAA....12.1197C}) is located. 

In this work, we present the TMTS system in section 2, including the enclosure, the mount, the multi-optical telescopes, the focuser, the photometric filters and the detector. In section 3 we present the software system and the survey strategy. The performance and initial scientific discoveries of the TMTS are presented in section 4. We summarize in section 5.

\section{Hardware System}

\subsection{Enclosure}

The TMTS system is located in an 4-m diameter clamshell dome with all open four movable shutters, which is driven by four independent motors. Mechanical limit switches are installed on each movable shutter to monitor opening or closing of the enclosure status, combined with the RS-232 protocol, we could control the dome and monitor the real-time status of the dome shutters remotely. Compared with the traditional follow-up rotation dome, this type of enclosure could reduce the failure rate during the observations, especially for the survey telescope with high frequency slew. All-open dome has the advantage of keeping the temperature in the interior of the dome close to the ambient temperature quickly, which could improve the stability of the local atmosphere and the local seeing.

\begin{figure*}[htbp]
\centering
\includegraphics[angle=0,scale=0.45]{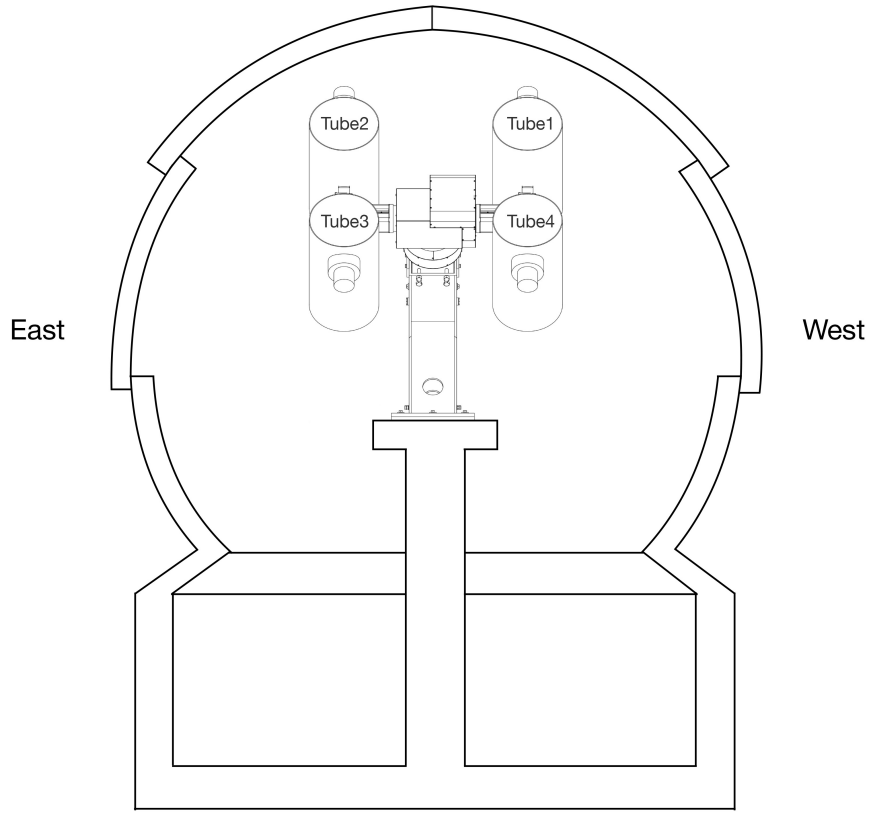}
\hfill \caption{Enclosure layout of the TMTS system with observing system inside; four tubes are arranged in the order marked in the plot.}
\label{fig1}
\end{figure*}

\begin{figure*}[htbp]
\centering
\includegraphics[angle=0,scale=0.1]{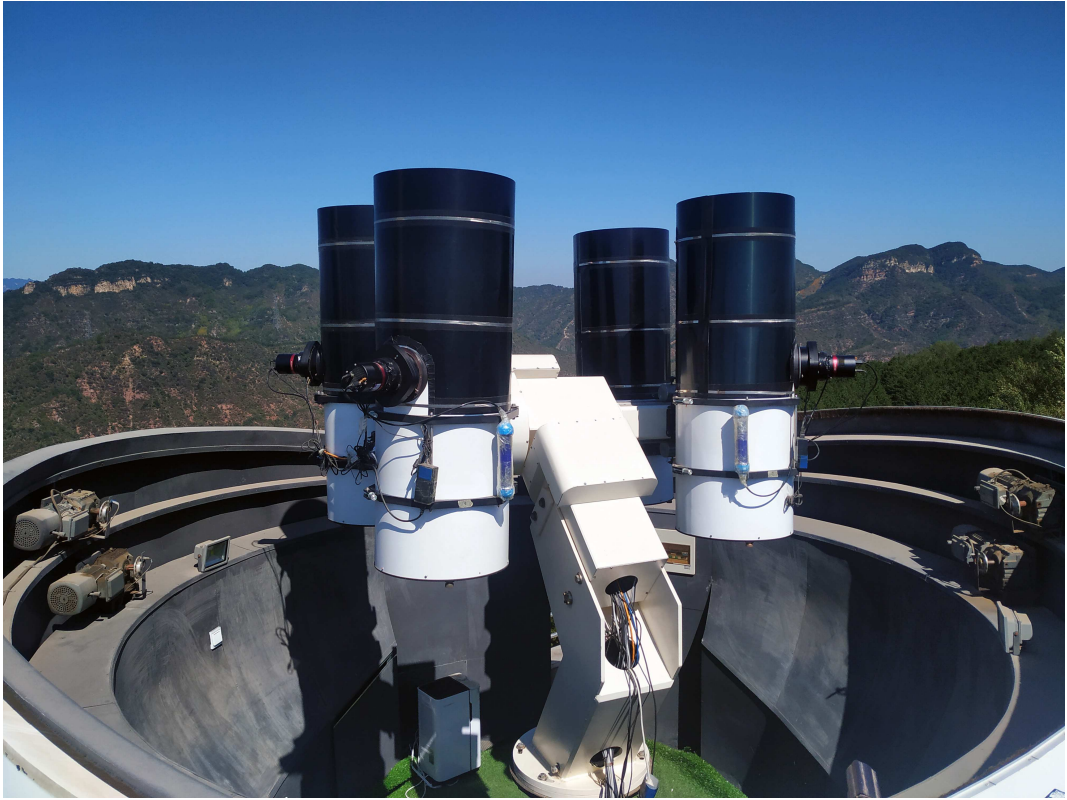}
\hfill \caption{Overview of the TMTS system inside the dome.}
\label{fig2}
\end{figure*}

\begin{figure*}[htbp]
\centering
\includegraphics[angle=0,scale=1.0]{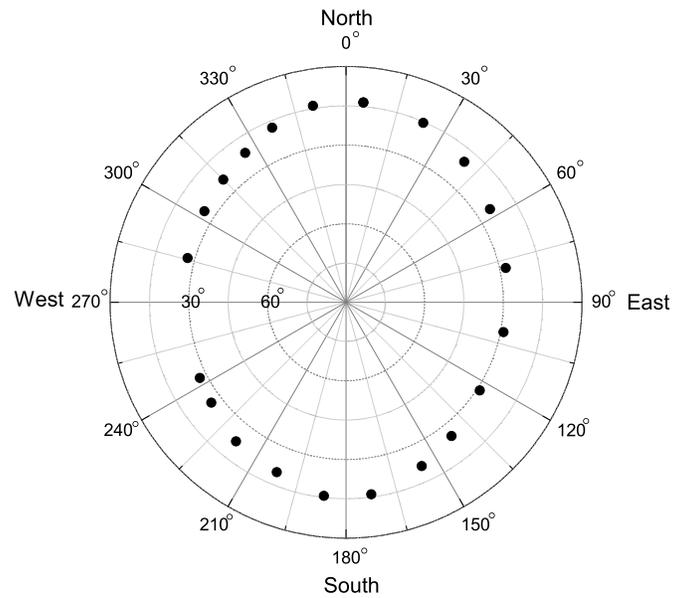}
\hfill \caption{Elevation pointing limit shown for the TMTS system. The black dots represent the actual elevation limitation for real observations, while the above regions show the observable area of the system.}
\label{fig3}
\end{figure*}

Enclosure layout of the TMTS system is shown in Figure~\ref{fig1}. Figure~\ref{fig2} presents the system of TMTS inside the dome. Considering the height of the mount and the bearing capacity analysis, we customized the pier for this system. We got the elevation limit curve for our observing system through actual telescope pointing observations, which could help us avoid the pointing limitation area and optimize the observation strategy. Elevation limit curve for the TMTS system is shown in Figure~\ref{fig3}, with the observable angles having a larger range in north and south compared to the east and west. The minimum pointing elevation angle in north and south could reach 15$^{\circ}$, and other two directions are about 30$^{\circ}$, which is related to the enclosure structural design and the opening mode of the dome shutters.

\subsection{Mount}

Classical mounts are mainly divided into two main categories: alt-azimuth mount and equatorial mount. These two subtypes of mount have both the advantages and disadvantages. Equatorial mount could avoid the problem of FoV rotation and the influence of integrity performance near the zenith, which is more suitable for small-aperture telescope system. As each of our telescope is about 50 kg and four telescopes are not very heavy for an equatorial mount, we chose the equatorial mount for our system. After systematic parameter analysis, we selected a German equatorial mount of MK 180 manufactured by ASTRO-TECHNIK\footnote{\url{https://www.astro-technik.com}} for our system.

\begin{figure*}[htbp]
\centering
\includegraphics[angle=0,scale=0.25]{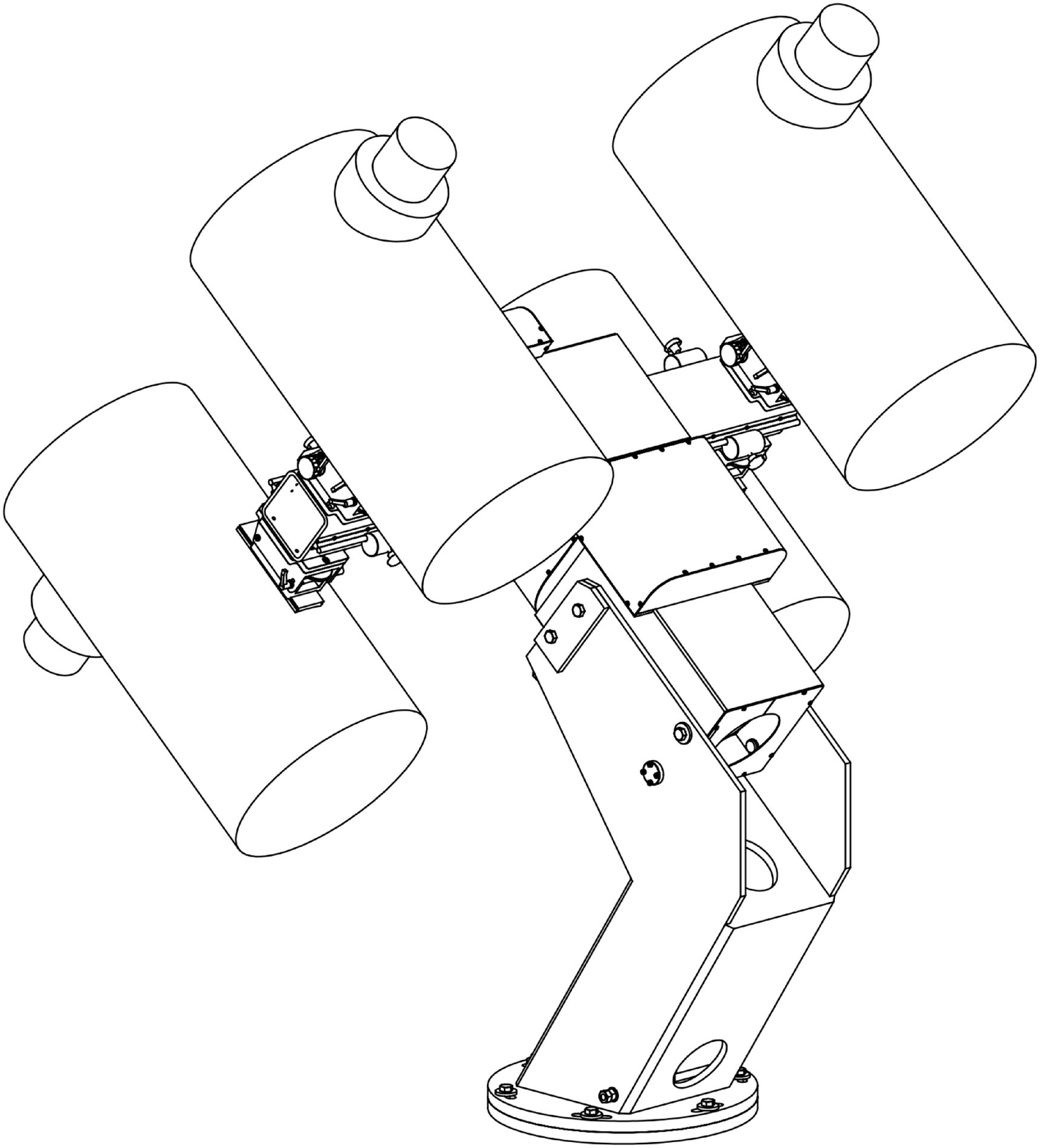}
\hfill \caption{Mechanical layout of four telescopes on Mount MK 180, which is offered and authorized by the manufacturer of ASTRO-TECHNIK. Pointing of each individual optical telescope could be adjusted with an angle range independently.}
\label{fig4}
\end{figure*}

\begin{table}[htbp]
\centering
\caption{Specifications of the Mount MK 180}
\begin{tabular}{ll}
\hline
\hline
Mount type	 		& German equatorial mount \\
Maximum payload		& 400 kg \\
Maximum slew rate	 	& $\textgreater$ 5$^{\circ}$ $s^{-1}$\\
Settling time			& $\textless$ 3 seconds\\
Pointing limitation		& Elevation $\textgreater$ 20 degrees\\
Slew	error				& $\textless$ 25 arcsec P-V \\
Tracking error			& $\textless$ 0.5 arcsec (peak to peak) with 3 minutes\\
\hline
\end{tabular}
\label{tab:mount}
\end{table}

Figure~\ref{fig4} shows the mechanical layout of four telescopes on the MK 180 mount, which has a 400 kg payload. As four telescopes are all installed on one mount, additional counterweights are not needed to balance and all the payload could be used on telescopes. Main parameters of the mount are listed in Table~\ref{tab:mount}. This mount used the industrial servo motors and the high resolution Renishaw resolute encoders. Maximum slew rate of this mount is faster than 5$^{\circ}$ $s^{-1}$, and the settling time is less than 3 seconds. This mount could reach all sky directions greater than 20 degrees above the horizon, which is suited for our survey. Mount tracking error for peak to peak is less than 0.5 arcsec without guiding for a duration of 3 minutes when observing sky area at 20 degrees above the horizon, and the slew error is within 25 arcsec P-V under the success pointing mode.

The TMTS system could have an angle range of $\pm$5 degrees around the central axis, allowing each individual optical telescope to point in two dimensional directions (Right ascension and declination) with independent stepless motion of the optical telescope being locked by screw mechanism. The adjusting precision is better than 3.0$\sim$4.0 arcminutes, and this special design allows different observation modes for the TMTS system.

\subsection{Telescope}

\begin{figure*}[htbp]
\centering
\includegraphics[angle=0,scale=0.3]{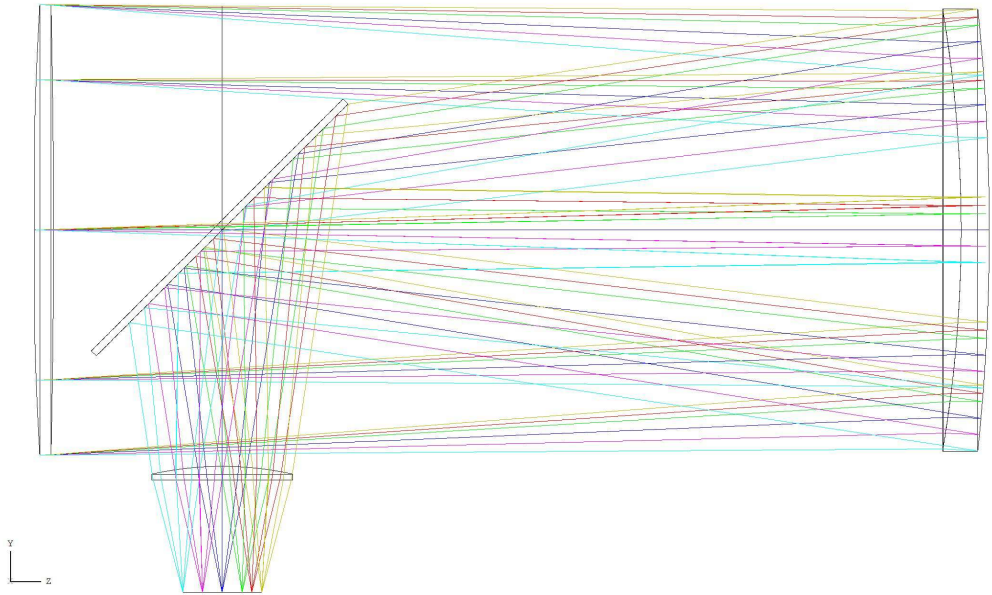}
\hfill \caption{Optical layout for the TMTS system, which is offered and authorized by the APM Telescopes.}
\label{fig5}
\end{figure*}

Telescope performance is determined by aperture, focal length, FoV, and image quality, etc. The TMTS system consists of four identical modified Hamilton-Newtonian astrographs, which are customized and offered by APM Telescopes\footnote{\url{https://www.professional-telescopes.com}}. Optical layout for the TMTS system is shown in Figure~\ref{fig5}. With the primary mirror placing in the closed tube and the third mirror near to the camera as corrector lens, the image quality will have a good correction.

Table~\ref{tab:telescope} presents the specifications of this Hamilton-Newtonian optical system. Each of the optical telescope has a diameter of 400 mm, made of Russian K8 material with 96\% coating for the reflection mirror. The secondary mirror has a short axis of 186 mm, with the central obscuration less than 46.5\%. The overall focal length is 1, 000 mm, with the $f$-ratio being 2.5. The optical telescopes are designed to have the bandpass covering from 400 $nm$ to 900 $nm$. Each telescope of the TMTS system has a focal plane of 80 mm in diameter (up to 100 mm). The corrected, non-vignetted field is 70 mm in diameter, which means that a large field of view can be realized when combining with a large-area detector.

\begin{table}[htbp]
\centering
\caption{Specifications of the Hamilton-Newtonian Optical System}
\begin{tabular}{ll}
\hline
\hline

Telescope diameter				& 400 mm \\
Working $f$-ratio				& $f$/2.5 \\
Focal length 					& 1, 000 mm \\
Secondary mirror size 			& 186 mm short axis\\
Central obscuration 				& $\textless$ 46.5$\%$ \\
Back focus distance 				& 105 mm\\
Corrected non-vignetted field 		& 70 mm in diameter\\
Focal plane 		 			& 80 mm (Up to 100 mm) in diameter\\
Encircled Energy 				& 85$\%$ within 0.6 arcsec\\
Designed working wavelength		& 400 $nm$$\sim$900 $nm$ \\
Mirror material 					& Russian K8 (like Schott BK-7)\\
Diffraction limited surface quality 	& Minimum of 1/4$\lambda$ P-V wavefront\\
Mirror coating reflectivity 			& 96$\%$\\
Designed RMS spotsize			& 1.76$\sim$3.46 micrometers \\
OTA weight					& $\sim$50 kg \\
\hline
\end{tabular}
\label{tab:telescope}
\end{table}

\subsection{Focuser}

For the TMTS system, we adopted the Atlas digital focuser of Finger Lakes Instrumentation (FLI)\footnote{\url{http://www.flicamera.com}} to adjust the focus of cameras. Specifications of the Atlas focuser is shown in table~\ref{tab:focuser}. The FLI Atlas focuser is a high load bearing focuser for large sensors, which could carry a maximum payload of 11.3 kg. This focuser has 105, 000 steps, with a resolution of 85 $nm$ per step. The micro-step controller is used for miniature stepper motor to adjust the precision steps. The custom linear bearings provide extreme torsional rigidity during the payload situation. Specially designed Zero Tilt Adapter$^{TM}$ help remove the effects of tilt, tip, or marred interfaces, resulting in final consistent performance in any orientation with the telescope position. Clear aperture of the focuser is 9.5 cm, allowing the whole light cone of the system pass through without being partially blocked at it current location in the light path.

\begin{table}[htbp]
\centering
\caption{Specifications of the Atlas Focuser}
\begin{tabular}{ll}
\hline
\hline
Clear aperture 			& 9.5 cm \\
Physical size			& 17.8$\times$17.8$\times$3.2 cm \\
Weight 				& 1.4 kg \\
Number of steps 		& 105, 000 \\
Steps resolution 		& 85 $nm$\\
Maximum payload		& 11.3 kg\\
Minimum payload		& 4.5 kg @ 15.2 cm \\
Travel distance			& 8.9 mm \\
Control protocol		& ASCOM\footnote{\url{https://ascom-standards.org}} compatible with auto focus capable \\
PC interface			& USB 2.0 with 12V DC power\\
Mechanical interface		& Zero-Tilt Adapter\\
\hline
\end{tabular}
\label{tab:focuser}
\end{table}

\subsection{Filters}

\begin{figure*}[htbp]
\centering
\includegraphics[angle=0,scale=1.0]{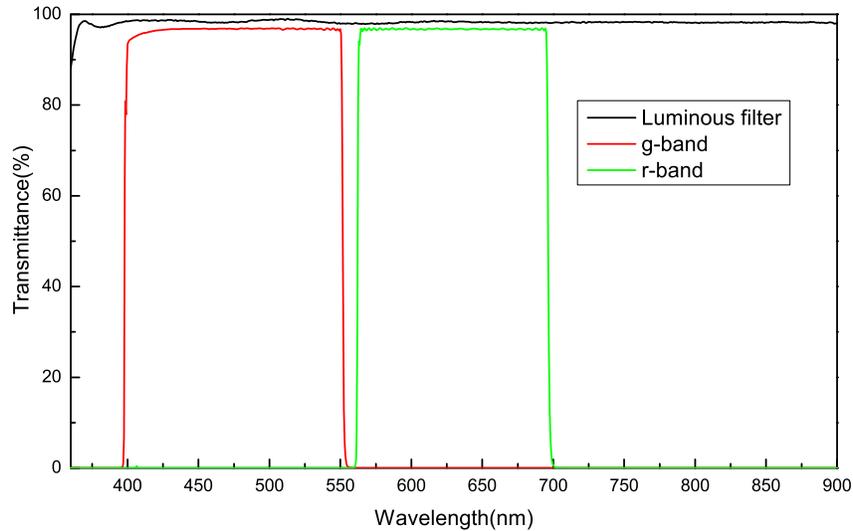}
\hfill \caption{Filters transmittance information for the TMTS system. The Luminous, $g$ and $r$ bands are marked by black, red and green lines, respectively.}
\label{fig6}
\end{figure*}

Considering the effective FoV for the TMTS system and the filter location in the light path, we customized the filters as 65 mm square and 5 mm thick. For regular survey, we use the Luminous filter, which has a wavelength coverage from 330 $nm$. For very high-cadence survey, we use the SDSS-like $g$ and $r$ filters, with $g$ filters installed on two telescopes and $r$ filters installed on the other two telescopes but pointing to the same sky areas as the former two. Central wavelength of the $g$ and $r$ filters are 4770 {\AA} and 6231 {\AA}, respectively, with the corresponding full width at half maximum (FWHM) being 1490 {\AA} and 1400 {\AA}. With such a combination of $g$- and $r$-band survey, we could get the $g-r$ color information for all of the survey targets. Figure~\ref{fig6} shows the filter information of the TMTS system.

\subsection{Detector}

As mean and median values of seeing at Xinglong Observatory of NAOC are around 1.9$''$ and 1.7$''$ \citep{2015PASP..127.1292Z}, respectively, combining the size of focal plane and the local typical seeing, we adopted the QHY 4040 detector for the TMTS system.  Considering the back focus length (i.e., 105 mm) for the TMTS system, the focuser, the filter holder, and the mechanical interfaces, the distance left for the detector is less than 32 mm. QHY 4040 detector has the back focal length of about 17 mm, which satisfied the  requirements. This detector is a monolithic CMOS with 4096$\times$4096 pixels, with a pixel size of 9.0 $u$m$\times$9.0 $u$m. This means an effective image area is about 36.9 mm$\times$36.9 mm, corresponding to a diagonal length of about 52.2 mm. This size is less than that of the focal plane for the TMTS system, indicating that only part of the FoV is used by QHY 4040.

\begin{figure*}[htbp]
\centering
\includegraphics[angle=0,scale=0.5]{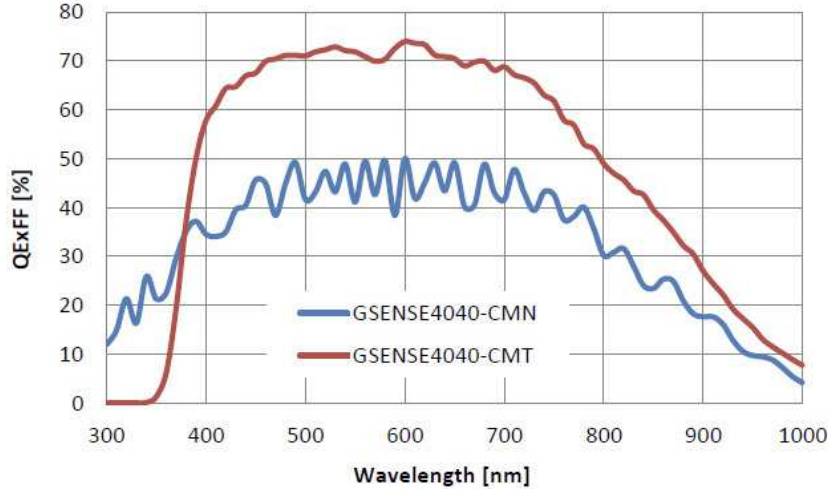}
\hfill \caption{QE of the QHY4040 Scientific CMOS detector, which is offered and authorized by the QHYCCD company. The red line of GSENSE4040-CMT is the sensor version used in our system.}
\label{fig7}
\end{figure*}

In order to improve the survey efficiency, read out speed should be as quickly as possible. The image from the QHY 4040 CMOS detector is about 32 MB, and the read out can be finished within 1 second, which is less than the slew time of the mount mentioned above. The typical read noise of QHY 4040 is about 4 e$^-$ at a gain of 32. QHY 4040 detector has 2-stage thermal electronic (TE) cooling that could reduce the sensor temperature to a value that is -35$^{\circ}$C below the ambient; and the dark current is about 0.05 e$^-$/pixel/sec at -15$^{\circ}$C. This implies that the TE cooling can reduce dark current significantly. Figure~\ref{fig7} shows the quantum efficiency (QE) of this camera, which is about 70\% at the wavelength from 400 $nm$ to 700 $nm$. This CMOS used the electric rolling shutter, which avoids the failures usually seen in traditional mechanical shutter. The surface glass above the CMOS sensor is a clear glass, while the glass of the CMOS chamber window is the AR+AR multiple layer coated for anti-reflection. This specific coating reduces the reflection and allows more light pass onto the sensor. Table~\ref{tab:CMOS} presents relevant parameters about the QHY4040 scientific CMOS.

\begin{table}[htbp]
\centering
\caption{Specifications of the QHY4040 Scientific CMOS}
\begin{tabular}{ll}
\hline
\hline
Mode					&QHY4040U3FSI\\
COMS sensor				& Gpixel Gense4040 FSI version\\
Effective pixel area			& 4096$\times$4096\\
Pixel size					& 9.0 $u$m$\times$ 9.0 $u$m\\
Effective pixels 				&16.8 Megapixels\\
Effective image area			& 36.9 mm$\times$36.9 mm\\
Cooling system 			& Dual stage TE cooler(-35$^{\circ}$C below ambient) \\
Readout noise 				& Typical 4 e$^-$ @ gain31 (16.5x)\\
Dark current 				& 0.05 e$^-$/pixel/sec @ -15$^{\circ}$C\\
Exposure time range 		& 20 $us$$\sim$600 sec\\
Sensor surface glass			& Clear glass\\
Chamber optic window glass	& AR+AR multiple layer coating anti-reflection glass\\
Shutter type 				& Electric rolling shutter \\
Fullwell 					& High gain channel $>$70 ke$^-$, low gain channel 26 ke$^-$\\
AD sample depth			& Dual 12 bit combine to 16 bit Image mode\\
Back focal length 			& 17.5($\pm$0.2) mm \\
\hline
\end{tabular}
\label{tab:CMOS}
\end{table}

\section{Software System and Survey Strategy}

\subsection{Software system}

The TMTS software system consists of two subsystems: observation subsystem and data reduction subsystem. Observation subsystem receives schedule lists as input, then it controls the telescope to point towards the targets and execute the exposure. Data reduction subsystem handles the pre-process of raw images, performing template subtraction and photometry, and searching transients. Softwares of observation subsystem are mainly hosted on Windows platform, data reduction subsystem are mainly operated on Linux platform.

\subsubsection{Observation subsystem}

Observation subsystem consists of softwares that control telescope, cameras, focusers and enclosure. We use FLI focuser\footnote{\url{http://www.flicamera.com}} drivers to control the focusers. Enclosure control is via ASCOM interface with iOptron Commander\footnote{\url{https://www.ioptron.com}}. Controls for focuser and enclosure are operated independently, while controls of the telescope and cameras are well integrated to allow for automatic observations. Software system for automatic observation can be divided into three parts: camera control, telescope control and observation central control. We introduce each part of our automatic observation system below.

Camera control software is written in Python language. As we have four cameras equipped on four individual telescopes, camera control software for each camera is on corresponding machine in order to ensure the speed and stability of data acquisition. This software is running within the script console of SharpCap\footnote{\url{https://www.sharpcap.co.uk}}, which is a GUI-based commercial astronomy camera capture software and could control the camera directly. When observation starts, the camera control software first connects to the camera, and it then listens on a TCP port and waits for the exposure command. Once it received the exposure command, the camera control invokes SharpCap API to start exposure and waits until the exposure finished. The camera control will set the flexible image transport system (FITS) keyword information to the images and save them with appropriate filenames. Finally the camera control will send a message to the observation central control to inform about the complete of the exposure command.

Telescope control software is operated by Sidereal Technology\footnote{\url{https://www.siderealtechnology.com}} through the servo controller. SiTechExe software has a GUI interface which is used for focusing and taking flat fields. It also provides a TCP interface to read the status of telescope and send out control commands. The observation central control system (OCCS) unitizes this feature to realize the automatic observations.

The OCCS is also written in Python language, and is operated on Linux platform. It receives a list of scheduling targets as input. For each target, it connects and sends the pointing command first to the telescope, while the control software will tell the telescope to point to the target. Once the telescope reaches the given coordinate position, the OCCS will connect and send exposure command to the control software of each camera, and it then waits until the camera controls return the exposure result. When receiving exposure results from all four cameras, the OCCS will move to the next target and repeat the above procedures.

When observation starts, the OCCS will calculate current position of moon using Python package skyfield\footnote{\url{https://pypi.org/project/skyfield}}, and use it throughout the night. Before pointing to one target, the OCCS will calculate the angular distance between the target and the moon; if the angular distance is too small, this target will be skipped. When the elevation of a target is too low, telescope control software will refuse to point and inform the caller; after informing the system, the OCCS will skip the target. With a well scheduled list of targets, the OCCS can perform automatic observations without human intervention.

\subsubsection{Data reduction subsystem}

\begin{figure*}[htbp]
\centering
\includegraphics[angle=0,scale=0.3]{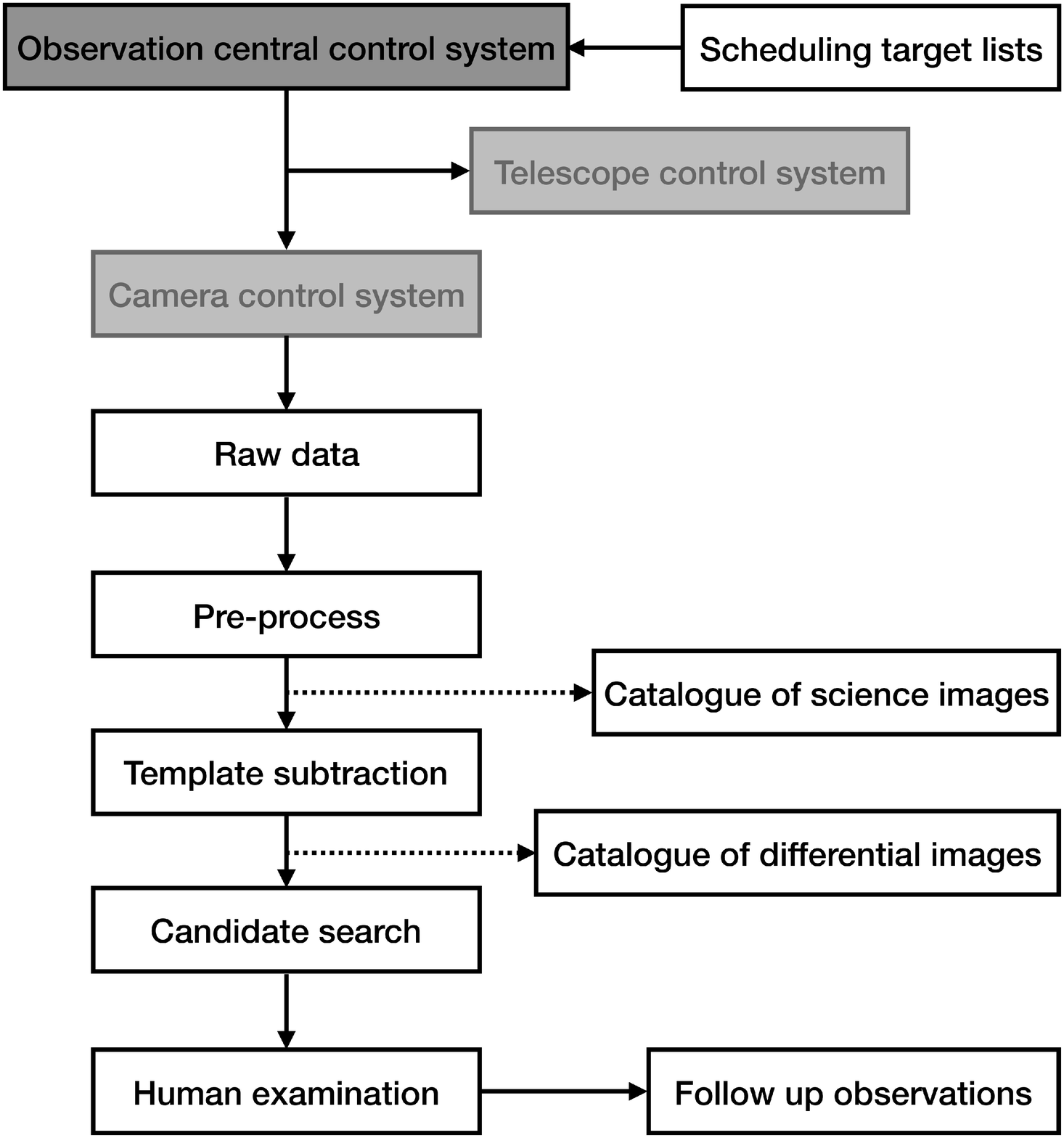}
\hfill \caption{A flowchart for the TMTS control system.}
\label{fig8}
\end{figure*}

The TMTS control system flowchart with observation and data reduction is shown in Figure~\ref{fig8}. We have modified the automatic pipeline used in TNTS \citep{2015RAA....15..215Z} to reduce the raw images from the TMTS observations. First, all raw images are calibrated by FITSH\footnote{\url{https://fitsh.net}}, which is mainly used for bias subtraction, dark subtraction, flat fielding and bad pixels masking. Then, we will use SCAMP \citep{2006ASPC..351..112B} to find an astrometric solution on calibrated images. Photometry for each image is performed by SExtractor \citep{1996A&AS..117..393B}, then the Luminous-band images will do the flux calibration by the $R$-band magnitudes from the PPM-Extended catalogue \citep{2008A&A...488..401R}; and the SDSS-like $g$ and $r$-band images are flux-calibrated by $g_{\mathrm{P1}}$ and $r_{\mathrm{P1}}$-band magnitudes from the Pan-STARRS catalogue\footnote{\url{https://catalogs.mast.stsci.edu/panstarrs}}, respectively.

Due to the low gain feature of the CMOS cameras, each exposure of the observation is splitted into multiple continuous frames to avoid saturation. Multiple frames of the same survey region are median combined into a combined image using SWarp\footnote{\url{http://www.astromatic.net/software/swarp}}\citep{2002ASPC..281..228B}. Then the combined images are subtracted by template images using a GPU sparse fast fourier transform (SFFT) template subtraction program. SExtractor is mainly used to extract candidates and measure corresponding photometric data from the subtracted images. After filtering by FWHM, flux radius, ellipticity and distance from nearby galaxy, each possible candidate will generate a triplet of crops of new, template and subtracted images, all these candidate informations will be linked to external database such as NASA/IPAC Extragalactic Database\footnote{\url{http://ned.ipac.caltech.edu}}, Transient Name Server\footnote{\url{https://wis-tns.weizmann.ac.il}} and IAU Minor Planet Center Minor Planet Checker\footnote{\url{https://minorplanetcenter.net/cgi-bin/checkmp.cgi}}. The triplet of crops, informations and links of candidates will be written into a webpage for further examinations of possible candidates. Data reduction pipeline could reduce multiple images up to a maximum of 16 images concurrently, this leads to nearly real-time reduction with only a few minutes delay, which is suitable for triggering quick follow-up observations.

\subsection{Survey Strategy}

\begin{figure*}[htbp]
\centering
\includegraphics[angle=0,scale=0.25]{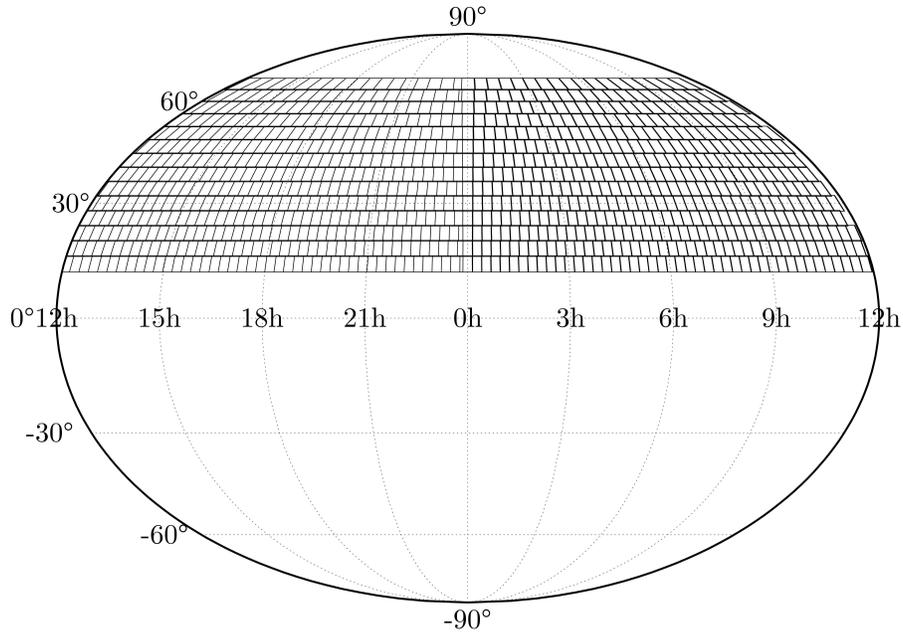}
\hfill \caption{The mosaic sky area monitored by the TMTS with declination ranging from 10$^{\circ}$ to 70$^{\circ}$ in the northern hemisphere. Each square represents about 17 deg$^2$.}
\label{fig9}
\end{figure*}

\begin{figure*}[htbp]
\centering
\includegraphics[angle=0,scale=0.35]{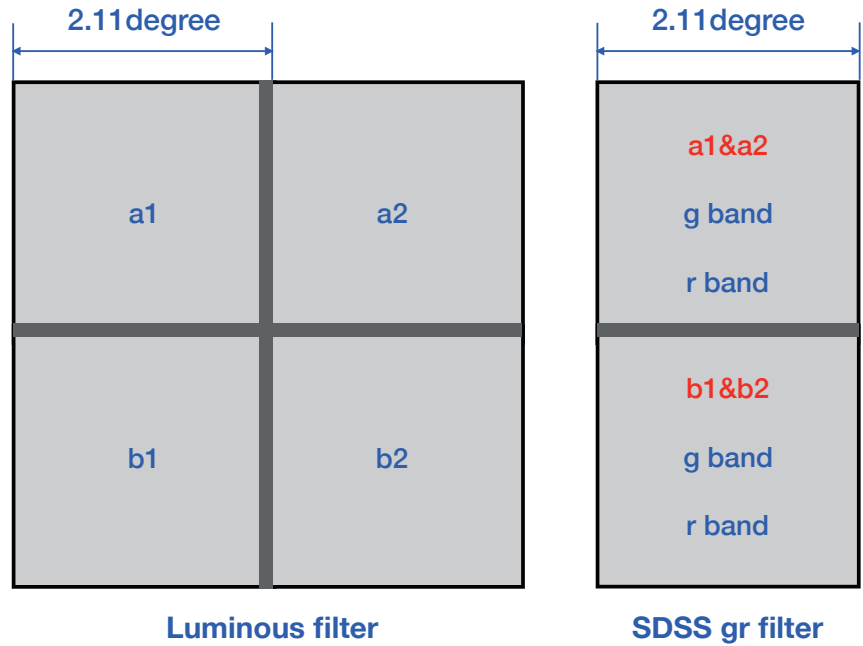}
\hfill \caption{Observation modes for the TMTS. Left: survey in monochromatic filter mode, with a mosaic FoV up to 17 deg$^2$. Right: survey in the combined g- and r-band mode, with a mosaic FoV about 8.5 deg$^2$.}
\label{fig10}
\end{figure*}

Supernova survey of the TMTS covers about 17 deg$^2$ sky regions from declination 10$^{\circ}$ to 70$^{\circ}$ in the north hemisphere. Figure~\ref{fig9} shows sky regions covered by the TMTS, each region is about 17 deg$^2$ with monochromatic filter mode.  Observation modes for the TMTS are shown in Figure~\ref{fig10}, where two modes of observations are used during the survey and could be adjusted through the screw mechanism of the mount.

Each sky region is designed to have a total exposure time of 60 seconds or 90 seconds, which is splitted into six continuous frames to avoid saturation. These six continuous frames will be combined into a single frame during data reduction, and the cosmic rays and tracks of satellites can be removed during this process. The scheduled observation plan for survey is semi-automatically generated everyday, and targets on sky regions are observed always near zenith, which could lead to a revisit cadence of nearly one day. The survey observation plan is specially arranged so that the adjoining sky regions could be observed successively, which will save the time for slewing the telescope. For 60 seconds exposure, the survey speed is about 42 regions/hour, which amounts to a sky area of about 7000 deg$^2$ per night in the Luminous filter mode during the winter observation season of Xinglong Observatory.

\section{TMTS Performance and Its Initial Scientific Results}

In this section, we focus on the performance of the TMTS system from the test survey conducted over the past few months. To get better image quality, we solved the aberration problem of the optical telescope and designed specific hood to avoid the effect of stray light. During the test survey conducted in Luminous filter, we discovered a lot of interesting candidates. Here we show some initial results and the detailed science results will be published in the forthcoming papers.

\subsection{Mount performance}

\begin{figure*}[htbp]
\centering
\includegraphics[angle=0,scale=1.0]{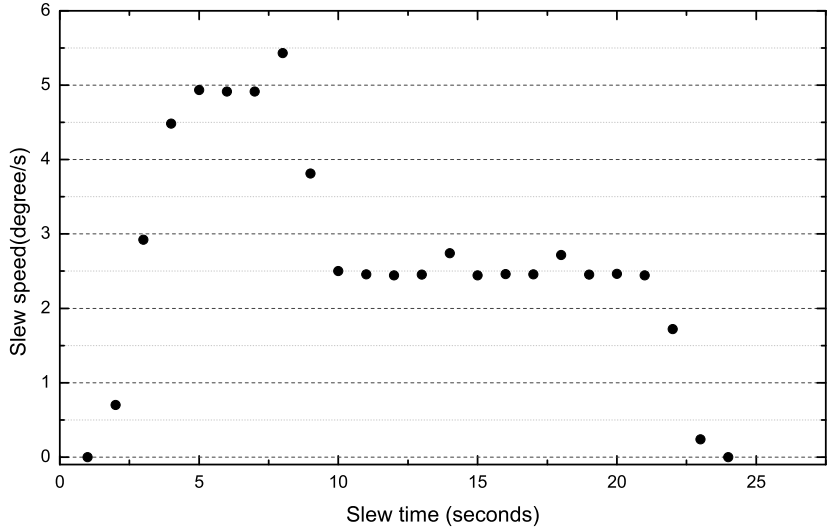}
\hfill \caption{An example of the mount slew speed measured for the TMTS system.}
\label{fig11}
\end{figure*}

Performance of an equatorial mount is usually evaluated by the payload, the slew speed, the pointing accuracy, and the tracking accuracy.  Mount slew status information could be read through the control software interface, such as tracking and slewing. To test the slew speed for our survey, we selected several stars with large angular distance, which will be conducive to check and monitor the mount motor status. As shown in Figure~\ref{fig11}, the maximum slew speed is larger than 5$^{\circ}$ $s^{-1}$, but drops to 2.5$^{\circ}$ $s^{-1}$ in the second half of the mount slewing process.

\begin{figure*}[htbp]
\centering
\includegraphics[angle=0,scale=0.3]{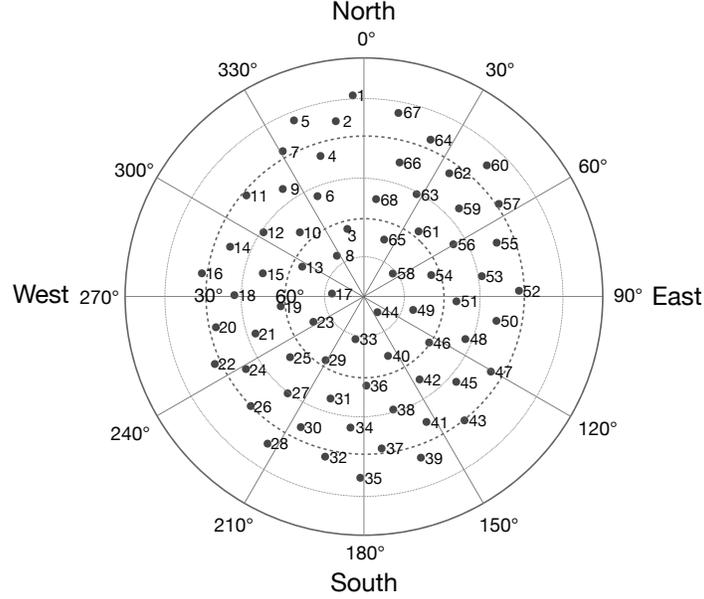}
\hfill \caption{Distributions of the pointing model data for the TMTS system. 68 stars are selected uniformly distributed in the sky; and differences between the actual position and the theoretical position of each star will be extracted for model fitting.}
\label{fig12}
\end{figure*}

\begin{figure*}[htbp]
\centering
\includegraphics[angle=0,scale=1.0]{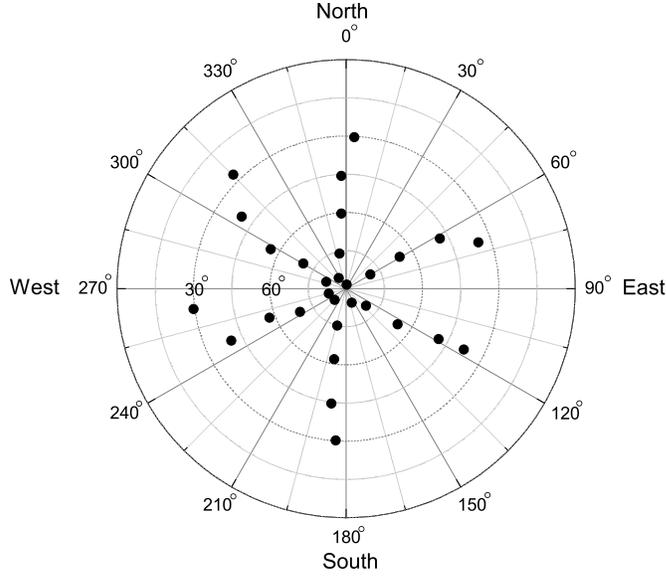}
\hfill \caption{Distributions of the actual pointing test derived for the TMTS system.}
\label{fig13}
\end{figure*}

We made the pointing model for the TMTS system after installation, using 68 stars uniformly distributed in the sky. Distributions of those selected stars are shown in Figure~\ref{fig12}. Considering the pointing limitation curve, stars in north and south directions are more than in west and east directions. Table~\ref{tab:pointing} shows the pointing data and the corresponding errors. After model fitting, the RMS pointing error is 14.9 arcsec, with 47.8 arcsec being the maximum. We also performed the actual pointing test after the pointing model fitting, as shown in Figure~\ref{fig13}. Table~\ref{tab:pointing_test} shows the detailed information about the pointing test, including the star's azimuth and elevation information, and the corresponding error data. Actual pointing test error is slightly larger than the pointing model, which is acceptable for a FoV of 2.11$^{\circ}$$\times$2.11$^{\circ}$.

We also tested the tracking performance of the mount during long-time exposure. This was done by extracting the star$'$s centroid coordinates on the CMOS chip, from which we could get the drifting value for all stars. After several experiments and calculations, we got the tracking error as 0.02$''$/second with the given pixel scale of 1.86$''$/pixel.

\begin{table}[htbp]
\scriptsize
\centering
\caption{Pointing Model Information for the TMTS System}
\begin{tabular}{lllllllllll}
\hline
Number&	Error	&Number	&Error	&Number	&Error	&Number	&Error	&Number	&Error	&	\\
	&arcseconds	&	&arcseconds	&	&arcseconds	&	&arcseconds	&	&arcseconds	&	\\
\hline
1	&	12.8	&	15	&	1.6	&	29	&	5.6	&	43	&	17.1	&	57	&	20.9	&	\\
2	&	16.6	&	16	&	6.1	&	30	&	3.6	&	44	&	30.0	&	58	&	10.2	&	\\
3	&	13.6	&	17	&	11.8	&	31	&	4.9	&	45	&	10.5	&	59	&	37.1	&	\\
4	&	12.0	&	18	&	9.1	&	32	&	1.8	&	46	&	15.6	&	60	&	35.4	&	\\
5	&	14.6	&	19	&	1.1	&	33	&	5.9	&	47	&	8.4	&	61	&	17.9	&	\\
6	&	9.1	&	20	&	10.4	&	34	&	10.4	&	48	&	6.1	&	62	&	10.3	&	\\
7	&	4.5	&	21	&	0.9	&	35	&	7.8	&	49	&	7.5	&	63	&	47.8	&	\\
8	&	10.8	&	22	&	5.6	&	36	&	6.8	&	50	&	35.2	&	64	&	12.7	&	\\
9	&	7.8	&	23	&	6.0	&	37	&	14.0	&	51	&	8.2	&	65	&	25.5	&	\\
10	&	6.5	&	24	&	18.7	&	38	&	13.2	&	52	&	5.1	&	66	&	22.6	&	\\
11	&	2.8	&	25	&	1.1	&	39	&	3.8	&	53	&	15.4	&	67	&	6.0	&	\\
12	&	6.9	&	26	&	7.5	&	40	&	21.5	&	54	&	5.4	&	68	&	7.4	&	\\
13	&	4.5	&	27	&	3.7	&	41	&	10.8	&	55	&	25.5	&		&		&	\\
14	&	4.3	&	28	&	16.1	&	42	&	2.0	&	56	&	2.1	&		&		&	\\
\hline
\end{tabular}
\label{tab:pointing}
\end{table}

\begin{table}[htbp]
\centering
\footnotesize
\caption{Measured Error Data of Pointing Test for the TMTS System}
\begin{tabular}{lllll}
\hline
Azimuth 	& 	Elevation	& 	Ra. error 	& Dec. error  \\
degree	&	degree	&	arcseconds	&arcseconds	\\
\hline
3.10		&	30.41	&	93.84	&	6.70	\\
325.09	&	84.91	&	1.91		&	46.04	\\
349.21	&	75.99	&	13.73	&	49.05	\\
356.38	&	60.40	&	33.43	&	22.71	\\
357.53	&	45.68	&	107.62	&	17.29	\\
\hline							
6.80		&	88.43	&	2.16		&	55.90	\\
59.16	&	78.98	&	2.92		&	78.44	\\
59.34	&	65.53	&	4.08		&	86.85	\\
61.76	&	48.28	&	6.54		&	106.29	\\
70.72	&	35.01	&	61.36	&	227.40	\\
\hline							
117.38	&	37.94	&	36.89	&	82.38	\\
118.61	&	48.64	&	26.28	&	55.08	\\
124.62	&	65.43	&	12.31	&	79.36	\\
130.38	&	79.70	&	9.11		&	76.77	\\
158.74	&	84.05	&	4.73		&	71.24	\\
\hline							
183.97	&	30.23	&	1.84		&	0.31		\\
187.39	&	44.56	&	3.16		&	25.35	\\
189.55	&	61.95	&	5.24		&	33.93	\\
193.63	&	75.11	&	5.35		&	51.92	\\
225.56	&	83.72	&	6.46		&	50.96	\\
\hline							
243.26	&	69.70	&	2.48		&	28.14	\\
245.63	&	40.45	&	14.76	&	20.64	\\
249.21	&	57.83	&	21.68	&	35.42	\\
254.19	&	82.94	&	5.80		&	44.79	\\
262.40	&	29.53	&	1.76		&	20.96	\\
\hline							
288.71	&	81.73	&	4.42		&	34.75	\\
297.67	&	56.62	&	27.82	&	36.66	\\
300.54	&	70.44	&	7.62		&	15.38	\\
304.59	&	40.19	&	0.30		&	11.49	\\
315.34	&	26.97	&	18.91	&	19.95	\\
\hline
\end{tabular}
\label{tab:pointing_test}
\end{table}

\subsection{Camera Performance}

\begin{figure*}[htbp]
\centering
\includegraphics[angle=0,scale=1.4]{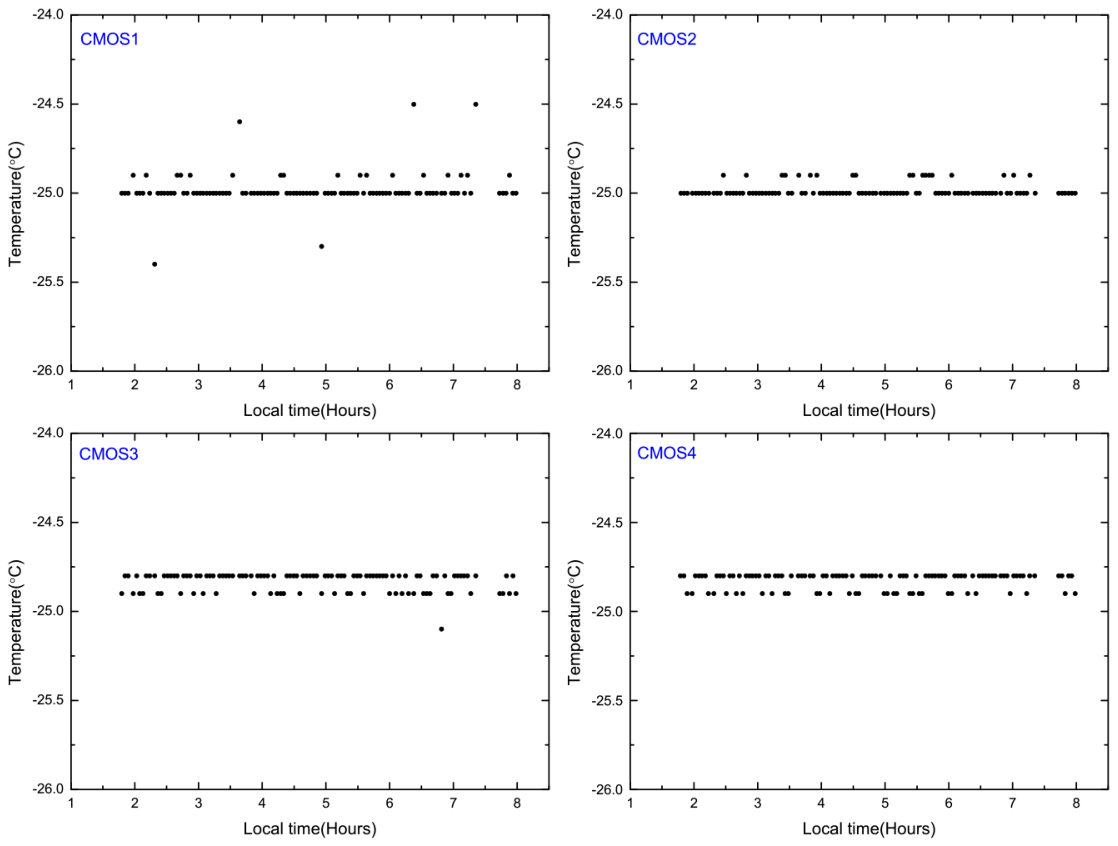}
\hfill \caption{Working temperature data of four CMOS collected in nearly 7 hours, the data are sampled every three minutes from the temperature sensor of CMOS chip.}
\label{fig14}
\end{figure*}

\begin{figure*}[htbp]
\centering
\includegraphics[angle=0,scale=1.0]{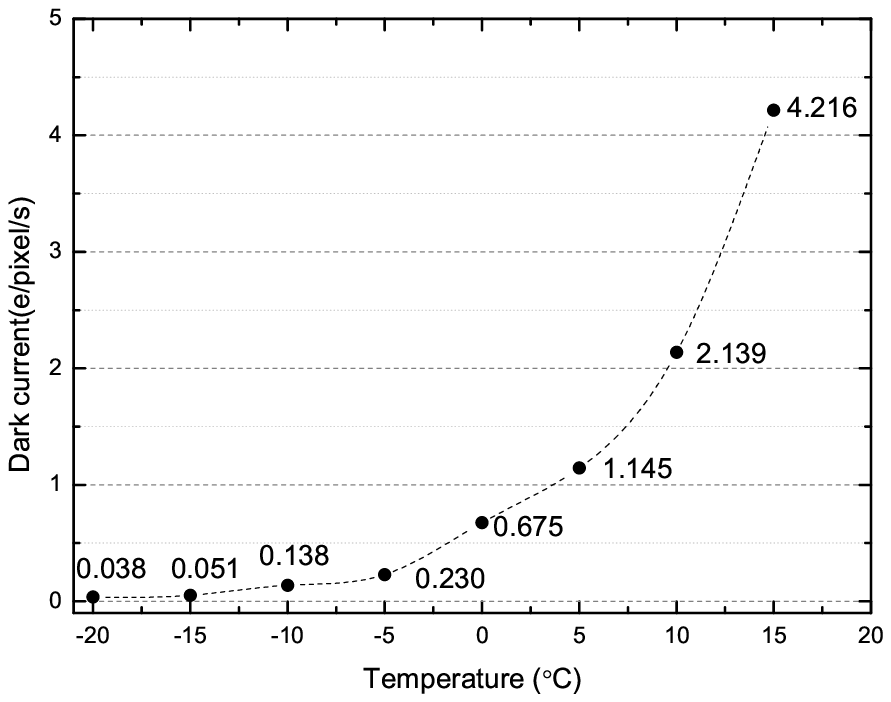}
\hfill \caption{Dark current of QHY 4040 scientific CMOS measured versus the chip working temperature of the chip.}
\label{fig15}
\end{figure*}

We further measured the stability of cooling temperature for the four CMOS detectors, as the working temperature directly influence the detector performance. Figure~\ref{fig14} shows the temperature change of each CMOS recorded in nearly 7 hours. All the temperature data are locked at $-$25$^{\circ}$C with an accuracy of 0.1$^{\circ}$C except for a few drift values. One can see that the temperature shows little change, ranging from -24.9$^{\circ}$C to -25$^{\circ}$C, suggesting that the working temperature is stable. As a result, the dark current shows little change as shown in Figure~\ref{fig15}.

\begin{figure*}[htbp]
\centering
\includegraphics[angle=0,scale=0.5]{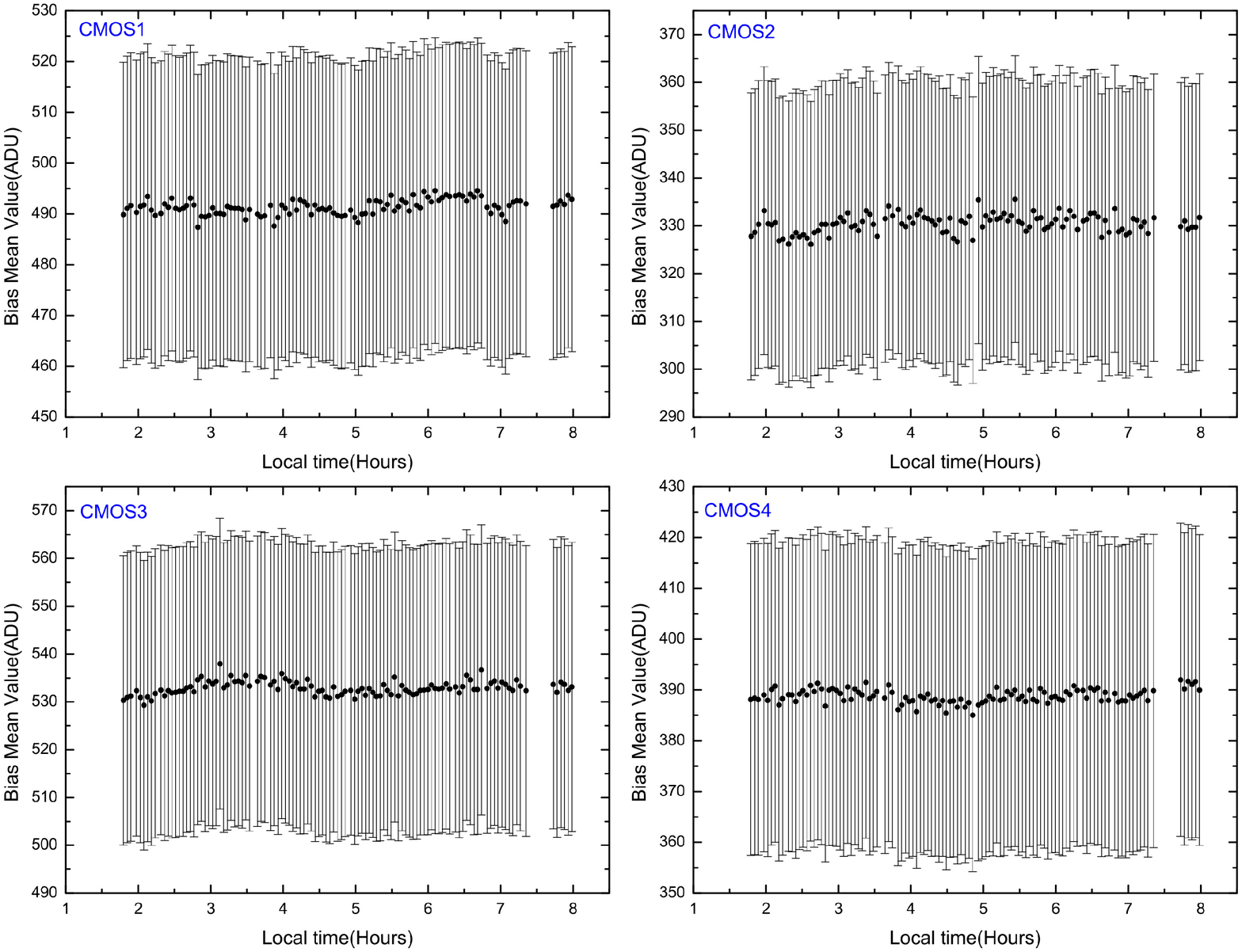}
\hfill \caption{Variations of the bias derived for four CMOS at -25$^{\circ}$C with the time duration of nearly 7 hours. Each error bar represents the STDEV of the corresponding bias value.}
\label{fig16}
\end{figure*}

We tested bias stability of CMOS on November 19, 2019, as the ambient temperature of Xinglong Observatory in this season was below 10$^{\circ}$C during the night. Considering the CMOS performance, bias test was set at a temperature of -25$^{\circ}$C. Figure~\ref{fig16} shows the bias test information of the four CMOS installed on the TMTS system. One can see that the bias level is different for different CMOS detectors, with a mean value and its corresponding STDEV being 491.3(1.52), 330.4(1.88), 532.9(1.50), 388.9(1.42) ADU, respectively. Different bias values are mainly due to differences in the bias voltage. The STDEV of these four CMOS is around 1.50.

\begin{figure*}[htbp]
\centering
\includegraphics[angle=0,scale=0.9]{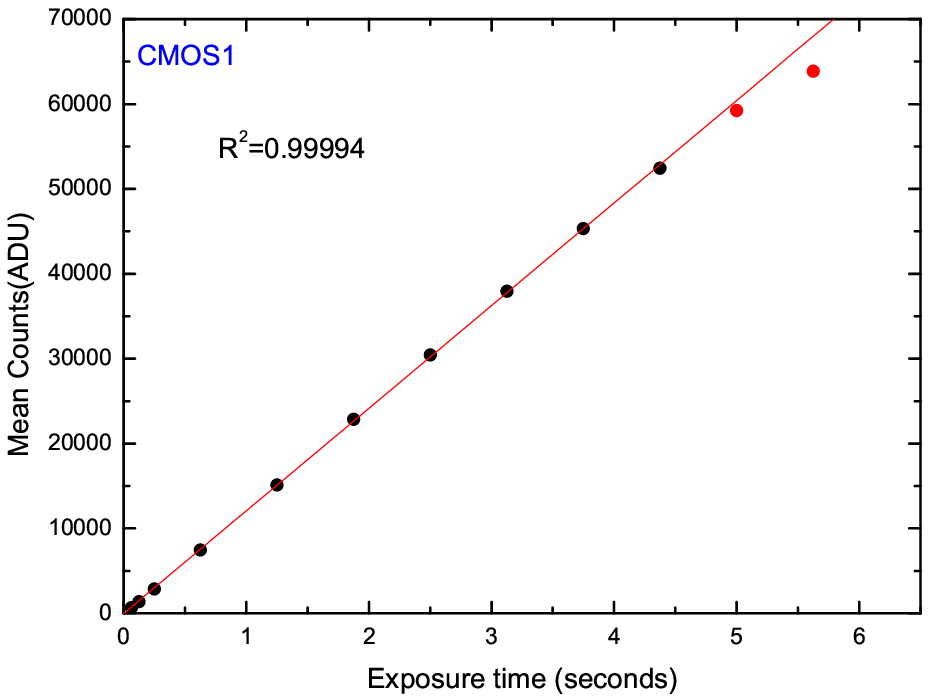}
\includegraphics[angle=0,scale=0.9]{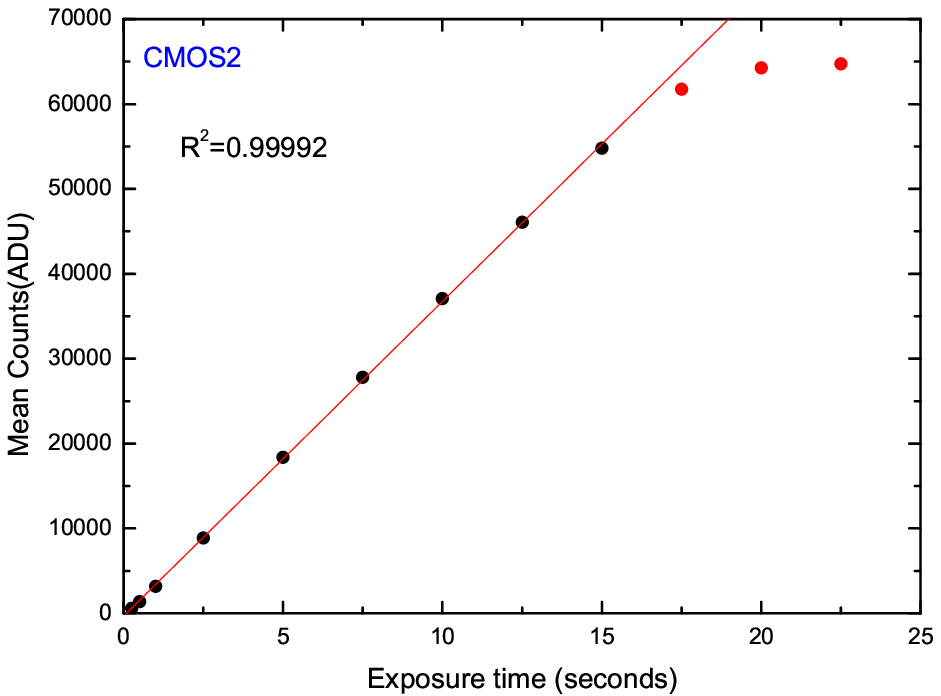}
\includegraphics[angle=0,scale=0.9]{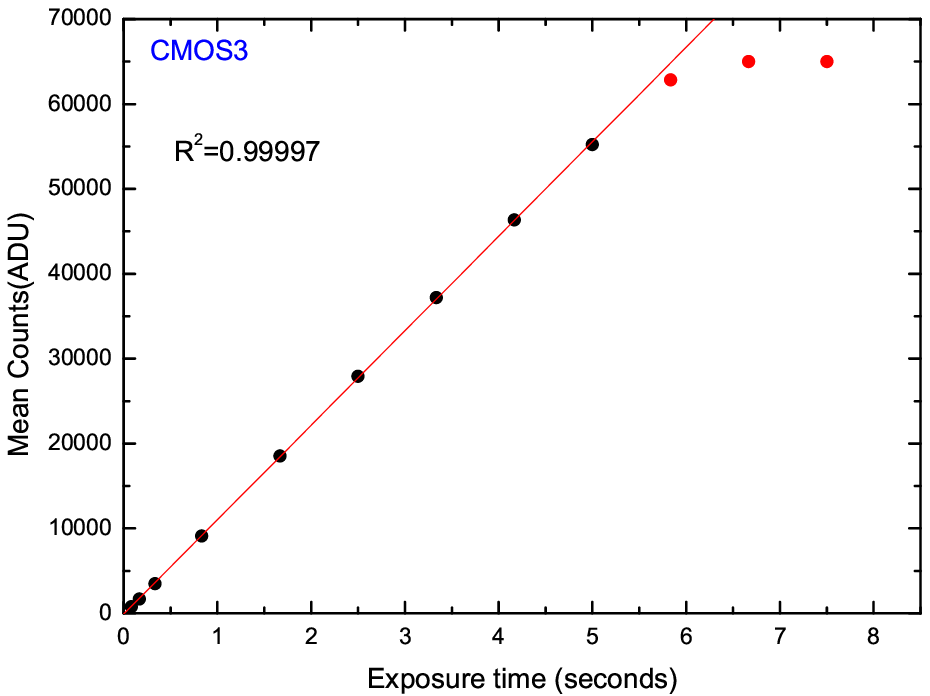}
\includegraphics[angle=0,scale=0.9]{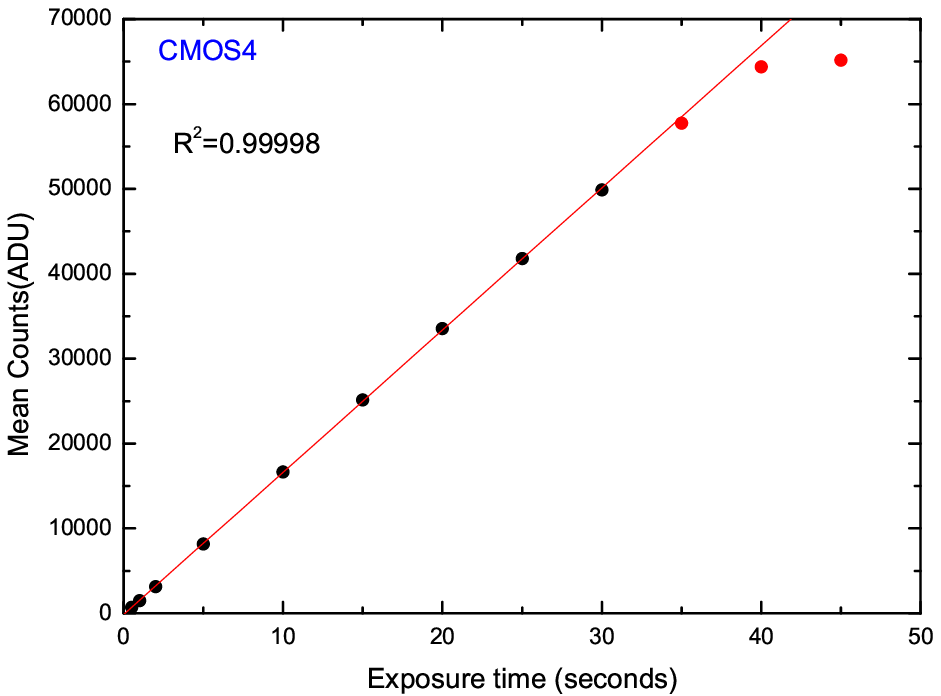}
\hfill \caption{Mean counts of four CMOS versus the exposure time for test of degree of linearity. The red line in each panel represents the linear fit to the black dots accuracy test, while the red dots show deviations from the linearity region perhaps due to saturation.}
\label{fig17}
\end{figure*}

The CMOS dark current was measured at a temperature of $-$25$^{\circ}$C, with CMOS1, CMOS2, CMOS3, and CMOS4 being 0.0740 e$^-$/pixel/s, 0.0253 e$^-$/pixel/s, 0.0314 e$^-$/pixel/s, and 0.0450 e$^-$/pixel/s, respectively. These values are very consistent with the given value of 0.05e$^-$/pixel/sec @$-$15$^{\circ}$C listed in Table~\ref{tab:CMOS}. The CMOS gain and readout noise measured at -25$^{\circ}$C are 0.40 e$^-$/ADU, 11.26 e$^-$ for CMOS1; 0.42 e$^-$/ADU, 11.82 e$^-$ for CMOS2; 0.39 e$^-$/ADU, 11.23 e$^-$ for CMOS3; 0.38 e$^-$/ADU, 11.09 e$^-$ for CMOS4, respectively. We finally selected the lowest gain mode for our observations, i.e., 0.40 e$^-$/ADU. We also tested the pixel linear correlation with different exposure time, and found that the linear correlation holds for the pixel value up to 55, 000 ADU, with the linear correlation coefficients being around 0.9999. Detailed information is shown in Figure~\ref{fig17}, where one can see that the pixel value curve tends to become flat above 60, 000 ADU.

\subsection{Optical Telescope Mechanical Modification}

\begin{figure*}[htbp]
\centering
\includegraphics[angle=0,scale=0.375]{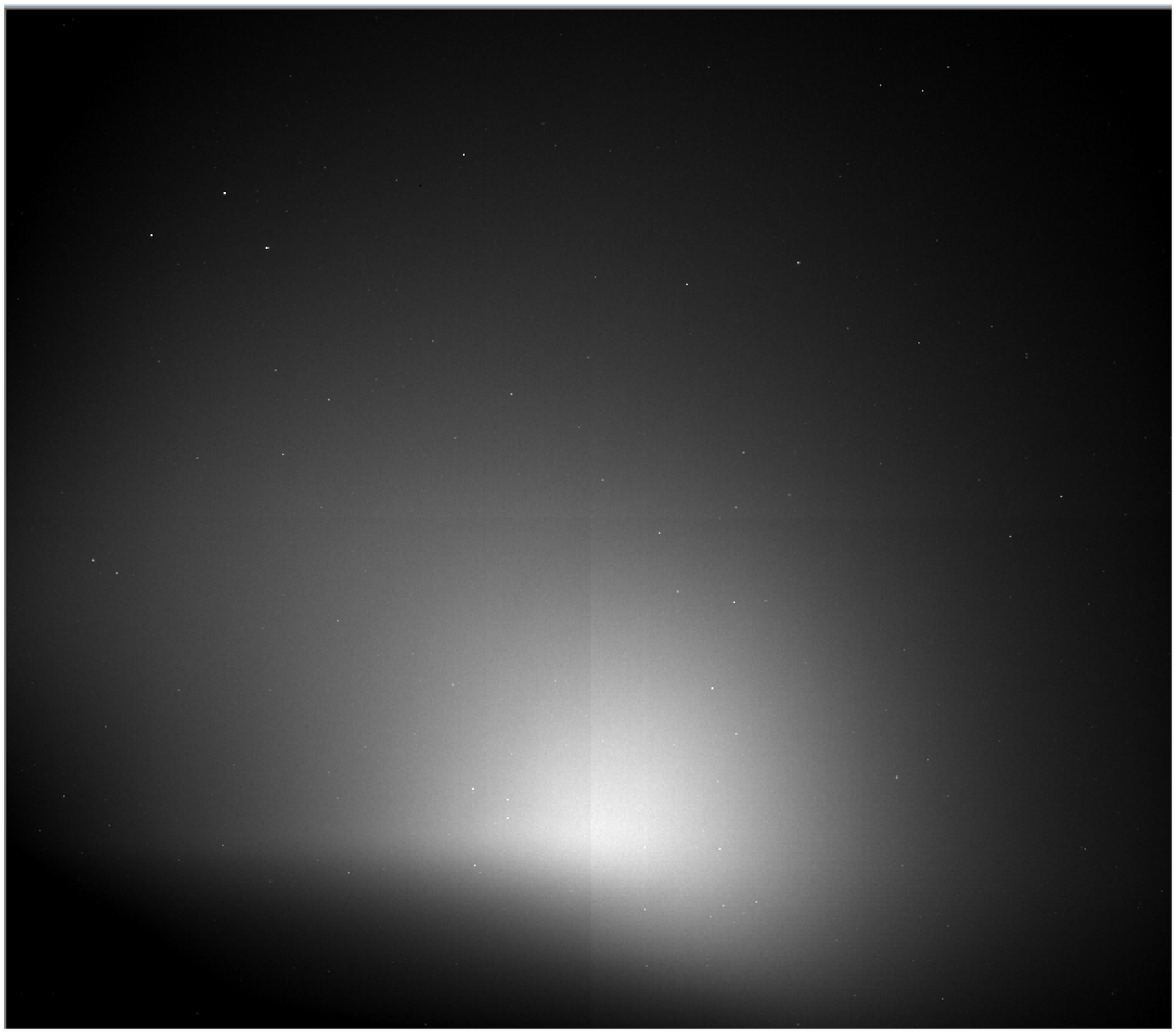}
\includegraphics[angle=0,scale=0.38]{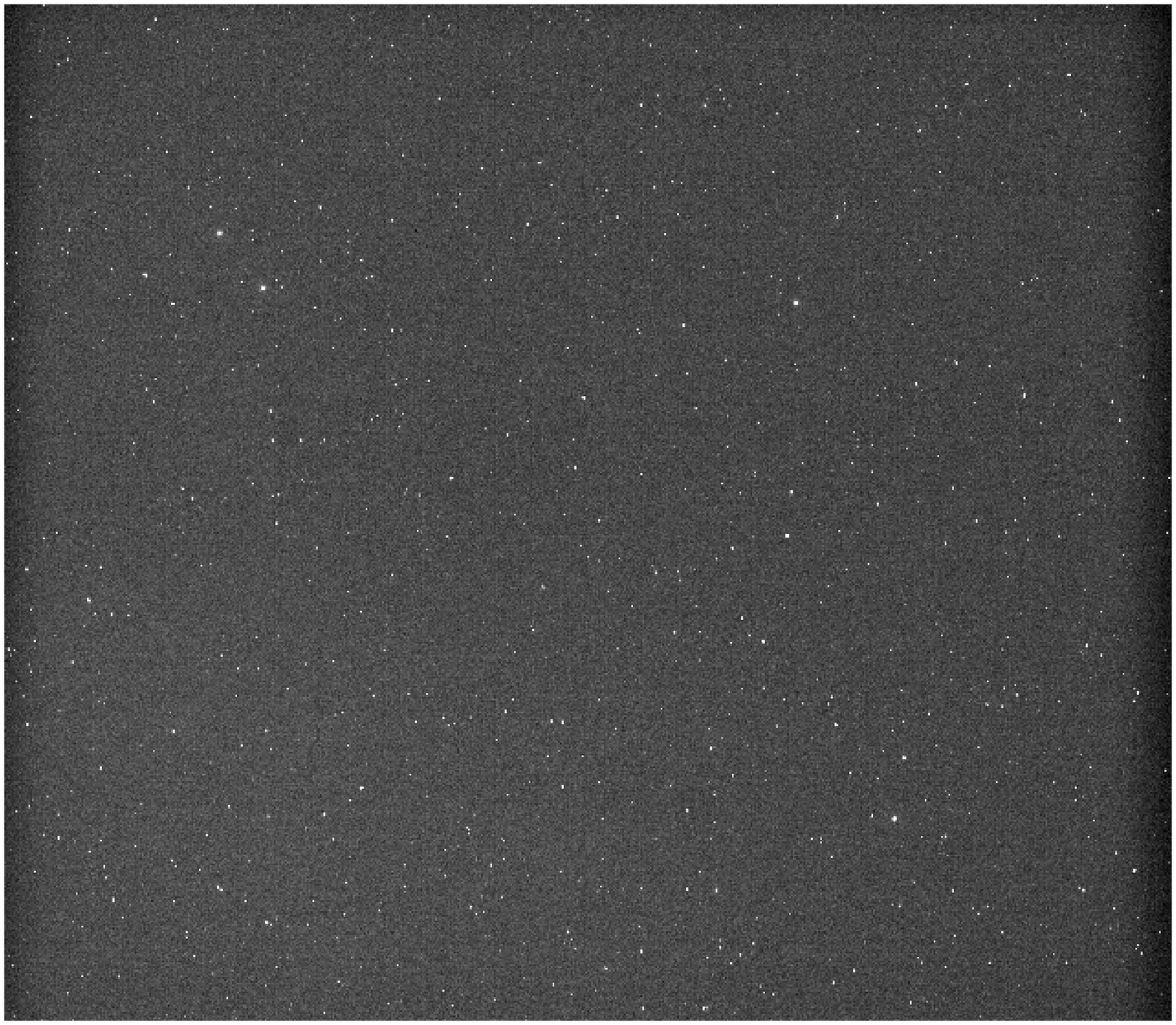}
\includegraphics[angle=0,scale=0.35]{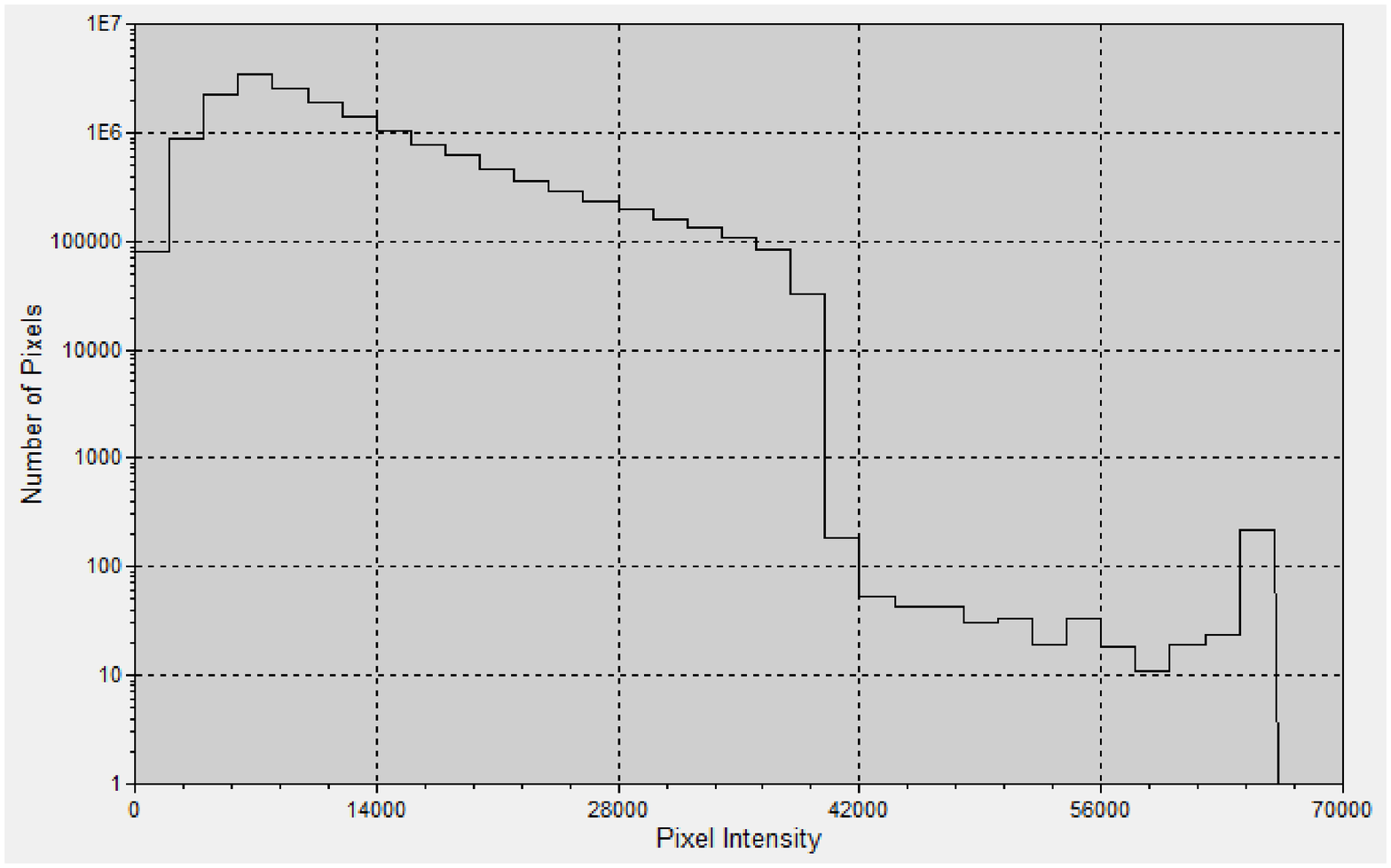}
\includegraphics[angle=0,scale=0.35]{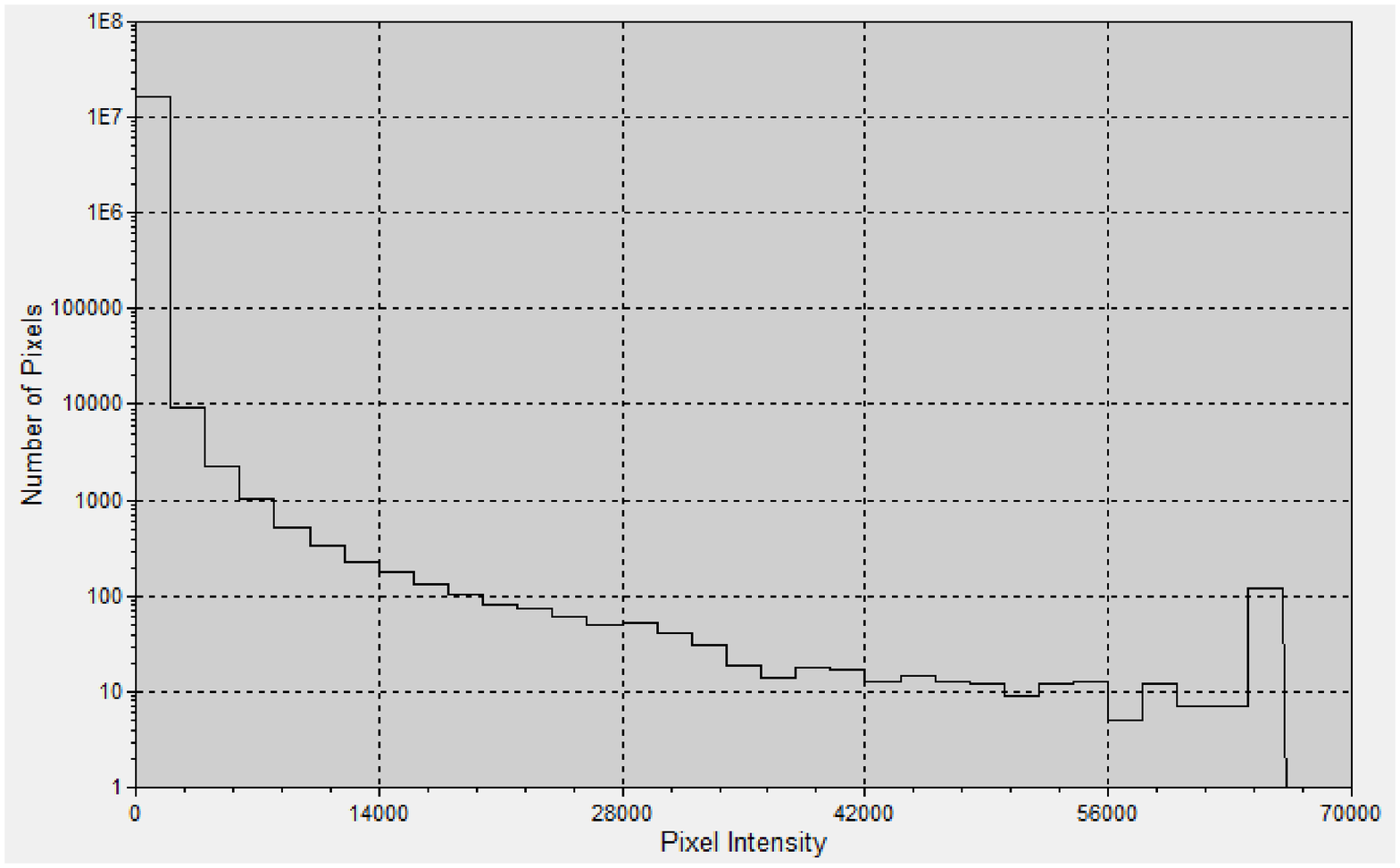}
\hfill \caption{Images taken by the TMTS system before and after the installations of optical telescope hood. Upper left: original image taken by the TMTS system during the night with large moon phase; Lower left: distribution of the pixel intensity of the image; Upper right: image taken by the TMTS system at the same position on the same night with the optical telescope-hood installed; Lower right: distribution of the corresponding pixel intensity for the image taken with hood.}
\label{fig18}
\end{figure*}

\begin{figure*}[htbp]
\centering
\includegraphics[angle=0,scale=0.5]{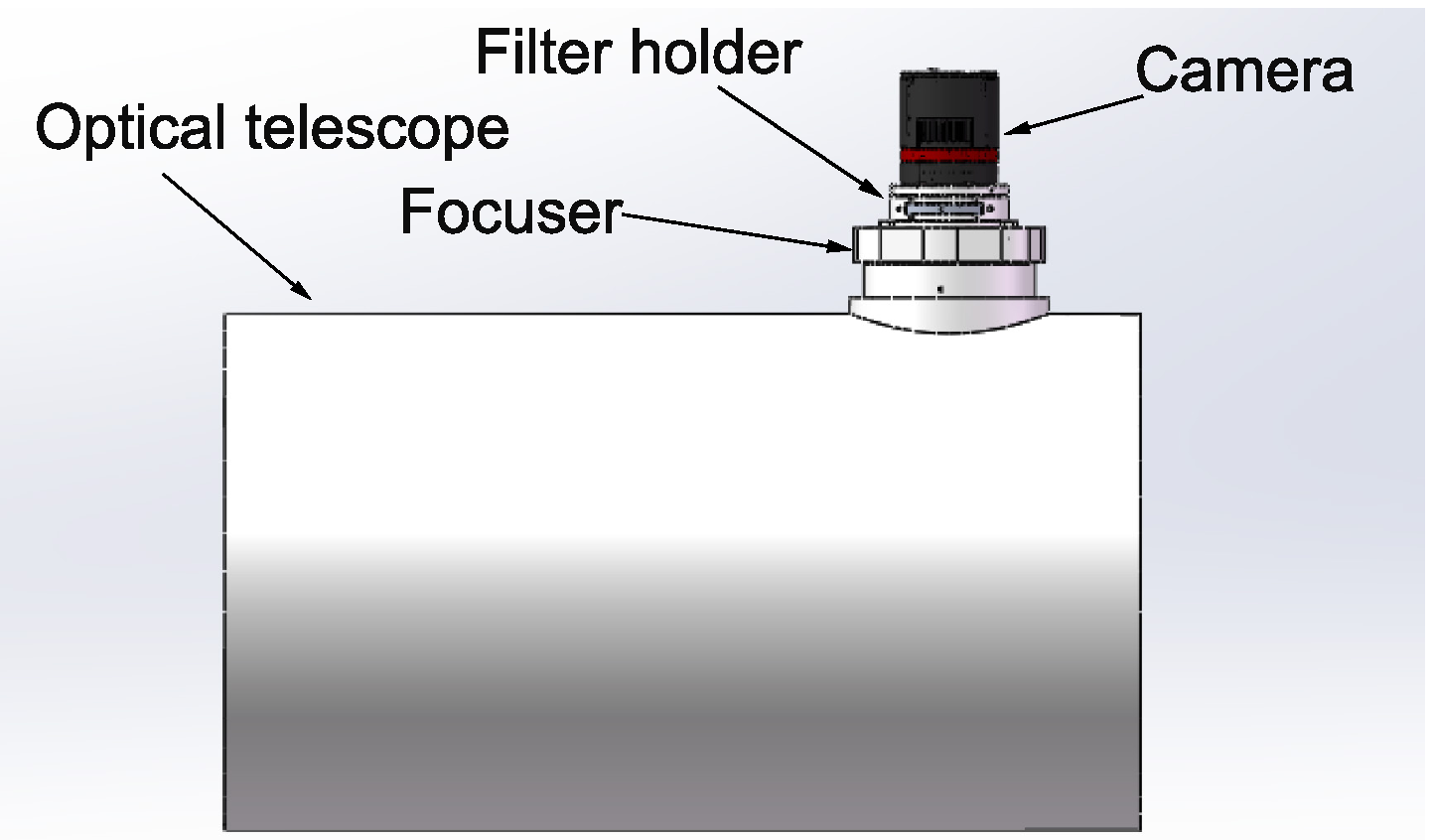}
\includegraphics[angle=0,scale=0.5]{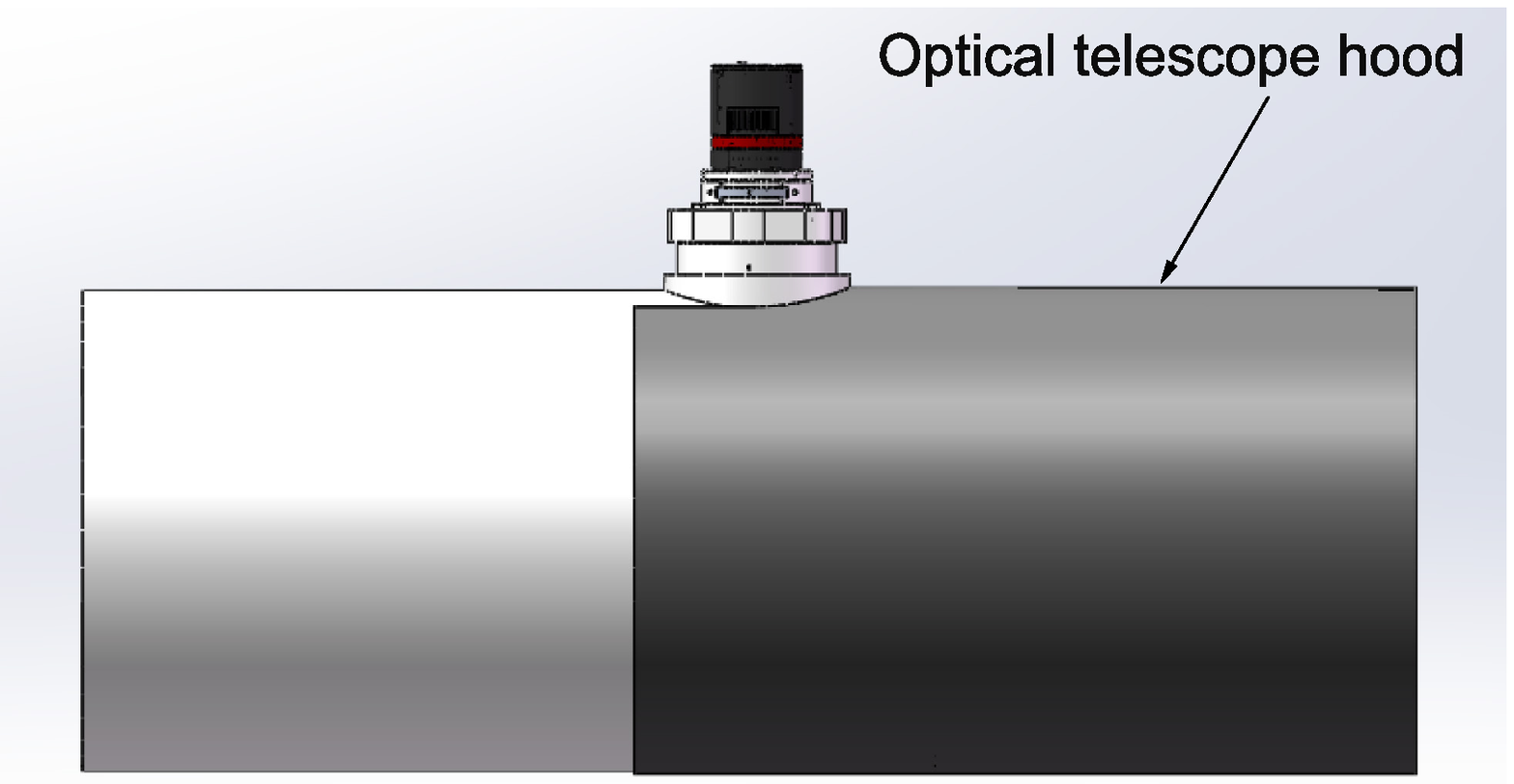}
\hfill \caption{Layout of the optical telescope for the TMTS system. Left: the assembly parts of the optical telescope, including focuser, filter holder and camera; Right: the optical telescope similar to the left side but with the hood installed in front.}
\label{fig19}
\end{figure*}

During the testing observations, we noticed that the stray light will be seen in the CMOS image at a certain angular distance between the telescope and the moon, especially during the nights with large moon phase, due to the optical design of this system. The original data influenced by the moonlight is shown on the left panel of Figure~\ref{fig18}, where one can see that a large light spot occupies a large proportion in the image. The pixel intensity in this affected region can be up to 40, 000 ADU, and the flux measurement of all the targets are seriously influenced by the stray light from the moon. To reduce such an effect, we designed a telescope-hood for our system, with the detailed layout information shown in Figure~\ref{fig19}. The hood is made by alloy material with special anodic oxidation treatment, which will minimize the light reflection in the optical light path. The right panel of Figure~\ref{fig18} shows the image taken after installing such a hood. Both of these two contrast images are taken with exposure of 1 second at the same position on the same night. With such a hood, the pixel intensity drops below 10, 000 ADU, indicating that such a hood is efficient in eliminating the stray light.

\begin{figure*}[htbp]
\centering
\includegraphics[angle=0,scale=1.0]{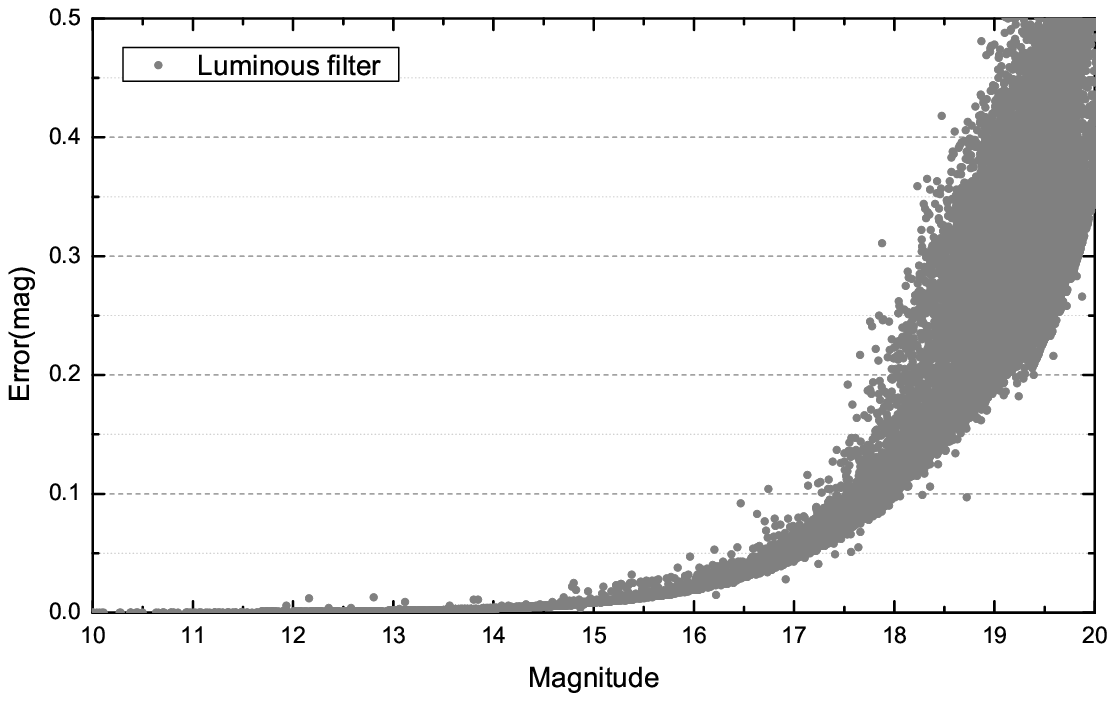}
\hfill \caption{Photometric performance of the TMTS system. The horizontal axis represents the detection limit measured from the Luminous-band image, while the vertical axis represents the corresponding errors in the measurements.}
\label{fig20}
\end{figure*}

Finally, we measured the limiting magnitudes for the TMTS system after solving the above problems encountered during the testing observations.  Figure~\ref{fig20} shows the limiting magnitude and its corresponding error, and all of the four telescopes could realize 3-$\sigma$ detection limit of 19.4 mag in Luminous filter for an exposure time of 60 seconds.

\subsection{Initial Scientific Results}

\begin{figure*}[htbp]
\centering
\includegraphics[angle=0,scale=0.45]{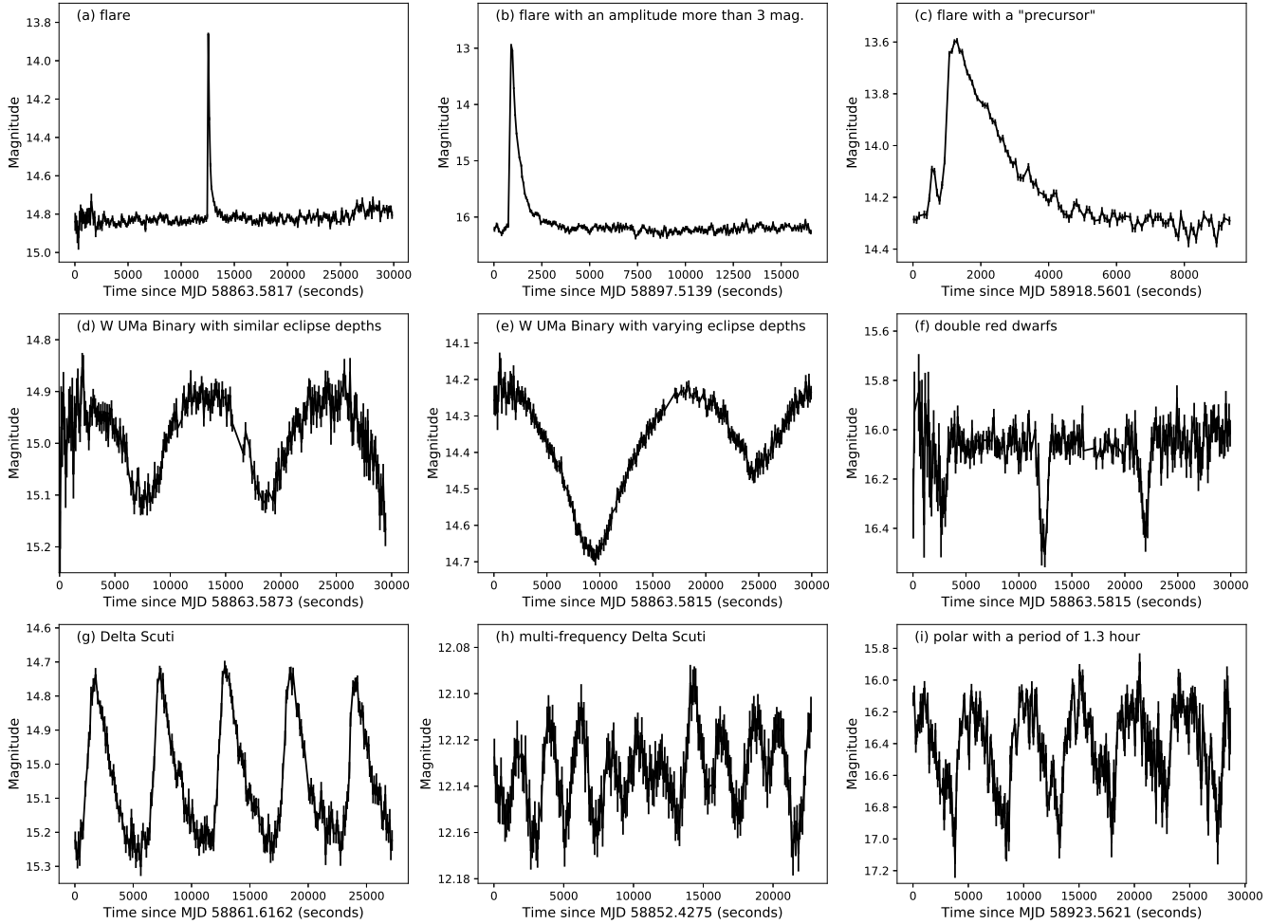}
\hfill \caption{Example light curves of transients detected by the TMTS system in Luminous band. Panels a-c present the discovered flare stars; Panels d and e present the binary stars with different eclipse depths; Panel f represents a double red dwarf system; Delta Scuti and polar stars are shown in panels g-h and panel i, respectively.}
\label{fig21}
\end{figure*}

During commissioning, the TMTS had scanned the LAMOST fields to get light curves for short-period variables or transients; these observations covered durations ranging from 3 hours to 10 hours with an exposure time of 10 seconds for each frame in Luminous filter. Many new astrophysical objects are discovered, including flare stars, eclipsing binaries, Delta Scuti variables, cataclysmic variable of AM Her type (polar), etc. Example light curves from the TMTS are shown in Figure~\ref{fig21}. We have also discovered the first supernova SN 2020esx \citep{2020TNSTR.849....1W} of the TMTS during the supernova survey, which is classified as a young type II supernova according to the spectral feature. Detailed scientific discoveries of the TMTS will be published in the forthcoming papers.

The high-cadence observations conducted by the TMTS allows us to reveal short-period variability (e.g. compact binaries) and resolve fast variability in light curves (e.g. flares). Through the Lomb-Scargle periodogram \citep{1982ApJ...263..835S}, tens of eclipsing binaries could be determined within one night. With the radial velocity information provided by the LAMOST spectra, the observations of the TMTS system can automatically estimate the lower limit for mass function and thus screen out those binaries with massive companions. These binaries are candidates of noninteracting black hole or neutron star \citep{2019Sci...366..637T}, which locate at an evolutionary stage lacking accretion and high energy radiation.

On the other hand, thanks to the high frame rate and wide FoV, the TMTS can capture the binaries with orbital period shorter than a few hours (i.e., ultracompact binaries), which are expected to be low-frequency gravitational-wave sources and will be detected by next-generation gravitational wave observatories (e.g. Tianqin and LISA, \cite{2009CQGra..26i4030N}; \cite{2012A&A...544A.153S}). Since the orbital period and mass of these binaries can be measured by electromagnetic observations, ultracompact binaries are thought to be $''$verification binaries$''$ for facilitating the functional tests of gravitational wave observations. Furthermore, the TMTS has successfully captured more than a dozen flare stars, which are rare, rapidly fading and unpredictable. Therefore, the TMTS also has the ability to discover other class of FOTs, the GRB afterglows. In theory, a part of GRBs is predicted to lack prompt high-energy emission due to strong beaming effects, but they may have afterglows in X-ray or optical bands (i.e., $''$orphan$''$ afterglows, \cite{1997ApJ...487L...1R}, \cite{2000ApJ...537..785D}). The wide-field optical survey has the opportunity to discover these afterglows independent of high-energy triggers, and contributes to the measurements of GRB rate and spatial distribution. Moreover, the TMTS system will also have great chances to localize the electromagnetic counterparts of the gravitational wave sources detected by the LIGO/VIRGO gravitational wave observatories in the next run in the mid of 2022.

\section{Summary}

Many wide-field optical surveys have been developed for time-domain astronomy over the past decade, however, most existing surveys could not detect the transients in different bands at the same time. To get timely color/temperature evolution of the transients during the survey, we developed the TMTS system, which also has an advantage of the geographic position allowing to coordinate with the spectroscopic survey conducted by the LAMOST.

We have described the hardware system, software system, survey strategy, performance and initial scientific results for the TMTS system. During commissioning, we have solved the problems encountered and fully tested the stability and reliability of the whole hardware system, including the performance of the enclosure, the mount, the telescope, the focuser, the filters and the detector. We have introduced the software system and the survey strategy, and tested the system performance. The TMTS system could not only collect the multiband information of the transients discovered during the survey, it can be also used for monochromatic survey to get a maximum FoV up to about 18 deg$^2$. For an exposure time of 60s, the TMTS observation could have a detection depth of $\sim$19.4 mag in Luminous filter and $\sim$18.7 mag in SDSS $r$ filter (3-$\sigma$). Numerous transients have been found during the first few months' testing observations, such as supernovae, flare stars, Delta Scutti stars, short-term eclising binary, and AGN etc; and the formal g- and r-band color survey will start from October 2020 when the color information for the discovered transients will be available by then. Moreover, it may also play an important role in searching light signals from the LIGO/VIRGO gravitational observatories in the future.

We acknowledge the kind support of the staffs from Xinglong Observatory of NAOC during the installation and commissioning of the TMTS system. This work is supported by the Ma Huateng Foundation and National Natural Science Foundation of China (NSFC grants 12033003, 11325313, 11633002, and 11761141001), and the National Program on Key Research and Development Project (grant no. 2016YFA0400803).

\end{document}